%% file: main.tex
\documentclass[sigconf]{acmart}

\usepackage{booktabs}
\usepackage{amsmath}
\usepackage{algorithmic}
\usepackage{graphicx}
\usepackage{textcomp}
\usepackage{xcolor}
\usepackage{xspace}
\usepackage{balance}
\usepackage{ifthen}
\usepackage{color}
\usepackage{upgreek}
\usepackage{caption}
\usepackage{subcaption}
\usepackage{soul}
\definecolor{todocolor}{RGB}{255,0,0} % Red color for TODO notes
\usepackage{tikz}
\usepackage{hyperref}

\setcopyright{none}
\renewcommand\footnotetextcopyrightpermission[1]{}
\newcommand{\ie}{\emph{i.e.,}\xspace}
\newcommand{\eg}{\emph{e.g.,}\xspace}

\newcommand{\etal}{\emph{et~al.}\xspace}

\newcommand{\figref}[1]{Fig.~\ref{#1}\xspace}

\newboolean{showcomments}
\setboolean{showcomments}{true}

\ifthenelse{\boolean{showcomments}}
  {\newcommand{\nb}[2]{
    \fbox{\bfseries\sffamily\scriptsize#1}
    {\sf\small$\blacktriangleright$\textit{#2}$\blacktriangleleft$}
   }
  }
  {\newcommand{\nb}[2]{}
  }

\newcommand\ccircle[2]{\tikz[baseline=(char.base)]{\node[circle, draw=black, fill=#1!60, minimum size=0.1cm, inner sep=2pt] (char) {\scriptsize #2};} }
\newcommand{\rpc}[1]{\texttt{\small{#1}}\xspace}

\newcommand{\toolname}{\textsc{vamp}\xspace}
\newcommand{\datasets}{33\xspace}

\newcommand{\SoBigDataITAck}{European Union - NextGenerationEU - National Recovery and Resilience Plan (Piano Nazionale di Ripresa e Resilienza, PNRR) - Project: “SoBigData.it - Strengthening the Italian RI for Social Mining and Big Data Analytics” - Prot. IR0000013 - Avviso n. 3264 del 28/12/2021\xspace}

\newcommand{\TAAck}{Territori Aperti (a project funded by Fondo Territori, Lavoro e Conoscenza CGIL CISL UIL)\xspace}

\begin{document}
\title{VAMP: Visual Analytics for Microservices Performance}

\author{Luca Traini}
%\authornote{Luca Traini is the corresponding author.}
\orcid{0000-0003-3676-0645}
\affiliation{%
  \institution{University of L'Aquila, Italy}
  \country{}
}
\email{luca.traini@univaq.it}

\author{Jessica Leone}
\orcid{0000-0002-2870-8161}
\affiliation{%
  \institution{University of L'Aquila, Italy}
  \country{}
}
\email{jessica.leone@student.univaq.it}

\author{Giovanni Stilo}
\orcid{0000-0002-2092-0213}
\affiliation{%
  \institution{University of L'Aquila, Italy}
  \country{}
}
\email{giovanni.stilo@univaq.it}

\author{Antinisca Di Marco}
\orcid{0000-0001-7214-9945}
\affiliation{%
  \institution{University of L'Aquila, Italy}
  \country{}
}
\email{antinisca.dimarco@univaq.it}

\titlenote{This is a pre-copy-editing, author-produced version of an article accepted for publication in \emph{The 39th ACM/SIGAPP Symposium on Applied Computing (SAC ’24)}.\\
%The final authenticated version is available online at: \url{https://doi.org/10.1145/3605098.3636069}.
Citation information: \href{https://doi.org/10.1145/3605098.3636069}{DOI 10.1145/3605098.3636069}}

%\renewcommand{\shortauthors}{J. Leone et al.}

\begin{abstract}
Analysis of microservices' performance is a considerably challenging task due to the multifaceted nature of these systems. Each request to a microservices system might raise several Remote Procedure Calls (RPCs) to services deployed on different servers and/or containers.
Existing distributed tracing tools leverage swimlane visualizations as the primary means to support performance analysis of microservices. These visualizations are particularly effective when it is needed to investigate individual end-to-end requests' performance behaviors. Still, they are substantially limited when more complex analyses are required, as when understanding the system-wide performance trends is needed.

To overcome this limitation, we introduce \toolname, an innovative visual analytics tool that enables, at once, the performance analysis of multiple end-to-end requests of a microservices system.
\toolname was built around the idea that having a wide set of interactive visualizations facilitates the analyses of the recurrent characteristics of requests and their relation w.r.t. the end-to-end performance behavior.
Through an evaluation of \datasets datasets from an established open-source microservices system, we demonstrate how \toolname aids in identifying RPC execution time deviations with significant impact on end-to-end performance.
Additionally, we show that \toolname can support in pinpointing meaningful structural patterns in end-to-end requests and their relationship with microservice performance behaviors.

%Video URL: \url{https://youtu.be/qMVOMt06EJE}
\end{abstract}




\maketitle

\input{introduction.tex}

\input{motivation.tex}

\input{tool.tex}

\input{evaluation}

\input{related}

\input{conclusion}

\begin{acks}
This work is partially supported by \SoBigDataITAck, and by \TAAck.

\end{acks}

\bibliographystyle{ACM-Reference-Format}
\bibliography{references.bib}

\end{document}

%% file: introduction.tex
%%%%%%%%%%%%%%%%%%%%%%%%%
%%%%%%%%%%%%%%%%%%%%%%%%%
\section{Introduction}
%%%%%%%%%%%%%%%%%%%%%%%%%
%%%%%%%%%%%%%%%%%%%%%%%%%

Microservices have emerged as a pivotal change in the software industry, paving the way to a novel paradigm for structuring the software development process.
This novel approach entails multiple independent teams responsible ``from development to deploy'' \cite{ohanlon2006} of loosely coupled independently deployable services~\cite{newman2015, ohanlon2006}. 
Due to their modular nature, microservices are particularly well-suited for the modern software industry, where rapidly releasing software updates and enhancements is a critical competitive advantage~\cite{rubin2016}.

Although beneficial in many aspects, microservices also introduce new challenges, especially when it comes to maintaining consistent software performance. This complexity arises from various elements.
 Firstly, the inherent complexity of these systems often hinders the adoption of proactive measures for performance assurance\cite{sridharan2017, veeraraghavan2016}, such as pre-production performance testing~\cite{jiang2015,laaber2018,traini2022c}.
Secondly, these proactive measures are often hampered by time and resource constraints due to the substantial pressure to deliver fast-to-market \cite{rubin2016, traini2022}.
Thirdly, microservices systems typically exhibit an emergent performance behavior in the field that is hard to predict in advance \cite{veeraraghavan2016}.
Finally, these systems undergo continuous software changes, with multiple releases occurring on a daily basis, and handle highly variable workloads \cite{ardelean2018}, which make them more vulnerable to unforeseen performance regressions~\cite{veeraraghavan2016,traini2021}.

These challenges have led to an increased interest in the concept of \emph{observability}~\cite{majors2022}, \ie the ability to have a holistic understanding of the system's performance by analyzing its logs, traces, and metrics.
Distributed tracing tools \cite{parker2020} are today widely used in practice to enhance observability of microservices systems \cite{mace2017}.
These tools track and record the propagation of requests as they flow through different RPCs and services of a microservices system \cite{sambasivan2016}, and provide visual aids to support performance analysis of end-to-end requests, \eg swimlane visualizations~\cite{Davidson2023, sigelman2010, jaeger}.

Despite their utility, distributed tracing tools have recently been criticized for their limited support for performance analysis~\cite{Davidson2023}. A common use case for these tools is the analysis of the system-wide performance behavior \cite{parker2020}, such as understanding the response time distributions of end-to-end requests \cite{Davidson2023}. However, current distributed tracing tools often fall short in this area, necessitating a switch between various visualization tools, which can make the process cumbersome and time-consuming~\cite{Davidson2023}. Indeed, they primarily focus on the analysis of individual requests, which has limited value unless it is compared with the performance behavior of the entire corpus of requests \cite{anand2020, parker2020, Davidson2023}.

In this paper, we introduce \toolname, an innovative visual analytics tool designed to enhance the performance analysis of microservices systems.
\toolname extends the conceptual proposal of Leone and Traini~\cite{Leone2023}.
The fundamental idea underpinning \toolname is to simplify the understanding of the relationship between request characteristics and end-to-end response time behavior through interactive charts and color-encoding techniques. \toolname comprises two main visualization components: an interactive tree that illustrates the workflow in terms of RPCs for multiple end-to-end requests, and an interactive histogram representing the end-to-end performance behavior of the requests under analysis.
Interaction with these visual components aids in identifying the unique characteristics of certain RPC execution paths with respect to some specific system performance behaviors.

We evaluate \toolname using \datasets datasets derived from TrainTicket~\cite{zhou2021}, an open-source microservices system widely utilized in previous software engineering research~\cite{traini2022b, zhang2022, li2022}.
Our findings demonstrate that \toolname enables the identification of notable and recurrent request characteristics associated with specific end-to-end response time behaviors.

A video of \toolname in action can be accessed at \url{https://youtu.be/qMVOMt06EJE}.

%% file: motivation.tex
%%%%%%%%%%%%%%%%%%%%%%%%%%
%%%%%%%%%%%%%%%%%%%%%%%%%%
\section{Motivation}
%%%%%%%%%%%%%%%%%%%%%%%%%%
%%%%%%%%%%%%%%%%%%%%%%%%%%

\begin{figure}[h]
         \centering
         \includegraphics[width=0.4\linewidth]{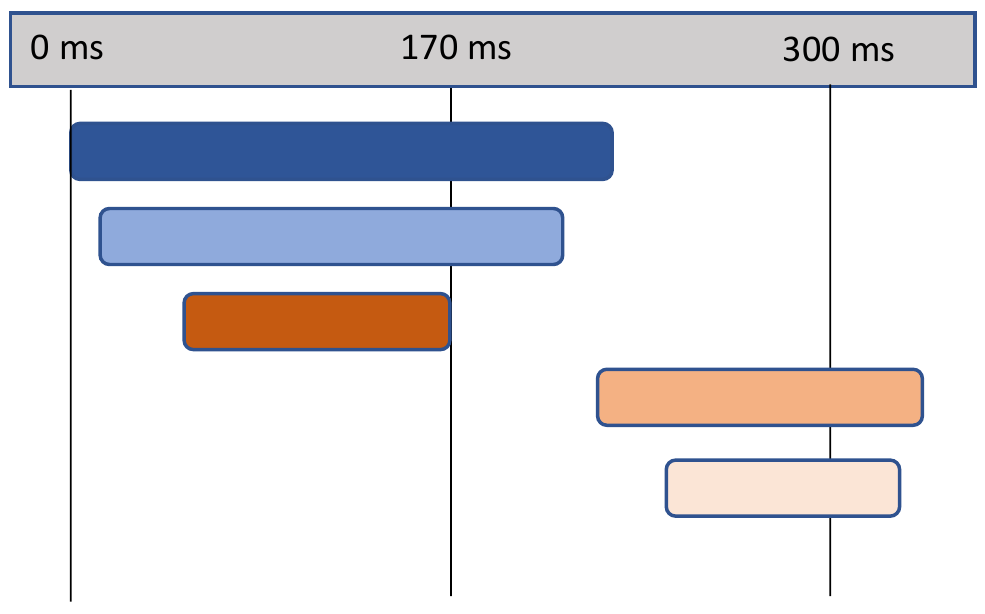}
         \caption{Swimlane visualization.}
         \label{fig:swimlane}
\end{figure}

The swimlane visualisation is the canonical way to visualize individual requests within distributed tracing tools \cite{kaldor2017, sigelman2010, jaeger, davidson2022}.
\figref{fig:swimlane} shows a representative example of a swimlane visualization. The visualization depicts a timeline of a single request, with RPCs depicted horizontally and sorted vertically to highlight their relationships.
This type of visualization proves highly beneficial for performance analysis of individual requests, allowing for a detailed investigation into how each RPC affects the overall end-to-end response time.

However, these visualizations exhibit certain limitations when it comes to conducting more complex performance analyses.
For instance, observing the performance of individual end-to-end requests might result in misleading insights if not contextualized appropriately~\cite{Davidson2023}. Indeed, a request's response time can only be deemed anomalous when compared with other requests of the same type \cite{anand2020}.
Additionally, engineers are often more inclined to investigate recurrent response time trends rather than focusing on the performance of individual requests \cite{parker2020}. Diverse end-to-end response time behaviors may be associated with specific request characteristics, such as particular RPC execution paths or RPC performance behaviors. Consequently, engineers may wish to identify these characteristics to uncover potential performance issues, and gather a more comprehensive picture of the system performance ~\cite{parker2020, krushevskaja2013, cortellessa2020}.

Distributed tracing tools currently lack sufficient support for this type of analysis, which often necessitates the concurrent use of multiple visualizations and tools, such as Jaeger~\cite{jaeger} and Kibana~\cite{kibana}~\cite{Davidson2023}. A naive strategy involves initially recognizing repetitive performance behaviors for further investigation, followed by the examination of individual requests to characterize relevant performance behaviors. This can be accomplished by detecting ``modes'' within the end-to-end response time distribution (for instance, using Kibana), which represent meaningful recurring performance behaviors. Following this, samples of requests associated with each mode can be extracted and examined individually (for instance, using Jaeger's swimlanes) to identify distinct characteristics that contribute to specific performance behaviors or modes.

However, this method can be particularly laborious as it requires manual inspection and comparison of multiple requests across diverse visualizations and tools. Moreover, even when the method is successful, it may not provide a satisfactory level of confidence.
Indeed, determining the specific characteristics associated with a particular distribution mode necessitates verifying that these characteristics appear \emph{exclusively} in requests that exhibit this particular end-to-end response time behavior.
This task can be challenging by using current tools.

\begin{figure}[h]
         \centering
         \includegraphics[width=0.6\linewidth]{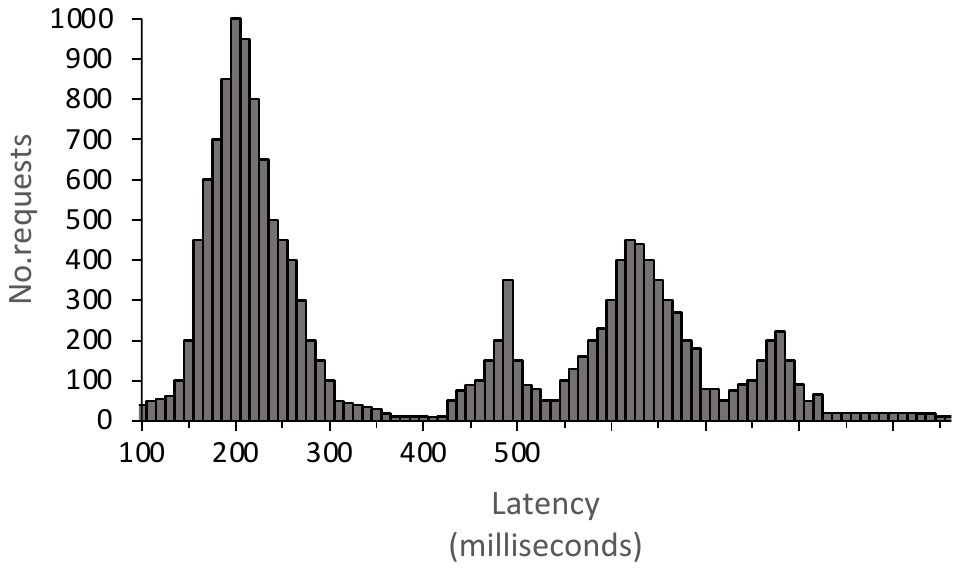}
         \caption{End-to-end response time distribution.}
         \label{fig:distribution}
\end{figure}

Consider the scenario illustrated in \figref{fig:distribution}, which represents the distribution of end-to-end response times for a specific type of request, such as loading a website homepage. As can be observed in the figure, requests demonstrate four distinct response time behaviors, \ie modes.
Suppose that the rightmost mode is characterized by a unique request characteristic, specifically an RPC that exhibits slower execution time\footnote{Henceforth, the term \emph{execution time} will be used to denote the response time of a generic RPC. Conversely, \emph{end-to-end response time} will refer to the response time of the root RPC that triggers all subsequent RPC invocations.} .
That is, this specific RPC shows increased execution time in all requests belonging to the rightmost mode (\eg due to an expensive task), but not in others.
With the current distributed tracing tools, identifying patterns like this can be particularly challenging.
Current distributed tracing tools lack targeted methods to simplify the analysis of RPC attributes, such as execution time, and their relationship with end-to-end response time.

%% file: tool.tex
%%%%%%%%%%%%%%%%%%%%%%%%%
%%%%%%%%%%%%%%%%%%%%%%%%%
\section{\toolname}
%%%%%%%%%%%%%%%%%%%%%%%%%
%%%%%%%%%%%%%%%%%%%%%%%%%

\toolname aims to enhance performance analysis of microservices systems by simplifying the investigation of attributes pertaining to specific RPC and their relationship with end-to-end response time.
In this section, we first introduce the core insights that underpin \toolname, its primary visual components, and the interaction modality. 
Then, we describe how these visual components fit within the \toolname dashboard, and detail the \toolname architecture and implementation.

\input{components}

\input{interaction}

\input{dashboard}

\input{implementation}

%% file: components.tex
\subsection{Visual Components}
The core insight behind \toolname is to make explicit the relationship between RPC attribute values and end-to-end response time.
RPC attributes could refer to several aspects, such as the \emph{frequency} of RPC invocation within a request, or the associated \emph{execution time}.
\toolname leverages two main interactive components to highlight this relationship: a \emph{tree} and a \emph{histogram}.
The tree provides an aggregated view of the requests' workflows in terms of RPC invocations, while the histogram displays a traditional distribution plot of the end-to-end response time.
Users can interact with the tree to examine how specific attribute values, related to a particular RPC execution path, influence the end-to-end response time; we refer to this as \emph{forward analysis}. Conversely, starting from the histogram, users can investigate how specific end-to-end response time behaviors are associated with certain RPC attribute values; this is referred to as \emph{backward analysis}.
In the following, we will first describe in detail the characteristics of these two main visual components. Then, in the subsequent subsection, we will detail the interaction modality of \toolname.

\begin{figure}
    \centering
     \begin{subfigure}[b]{0.3\textwidth}
         \centering
         \includegraphics[width=\textwidth]{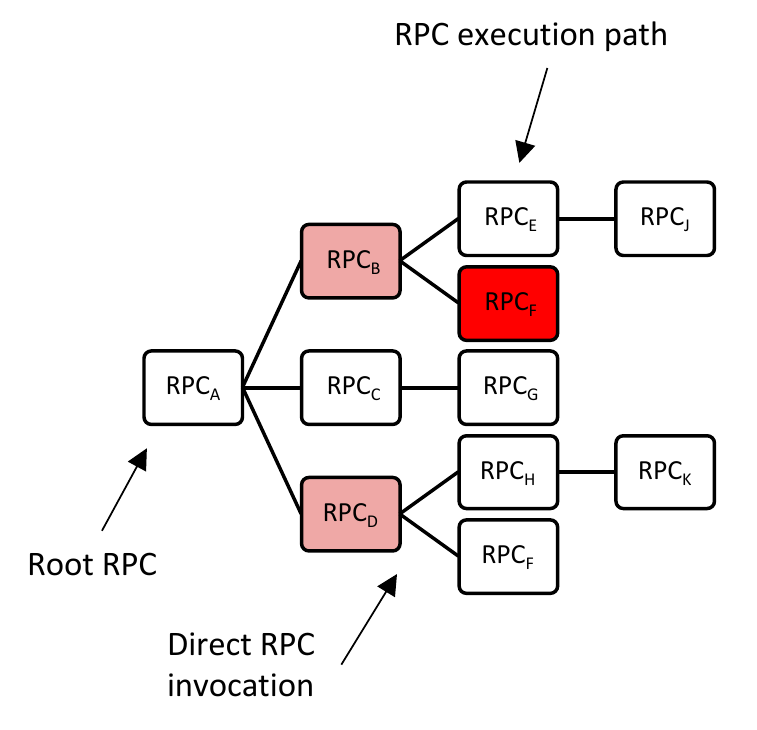}
         \caption{Tree}
         \label{fig:tree}
     \end{subfigure}
     \hfill
     \begin{subfigure}[b]{0.3\textwidth}
         \centering
         \includegraphics[width=\textwidth]{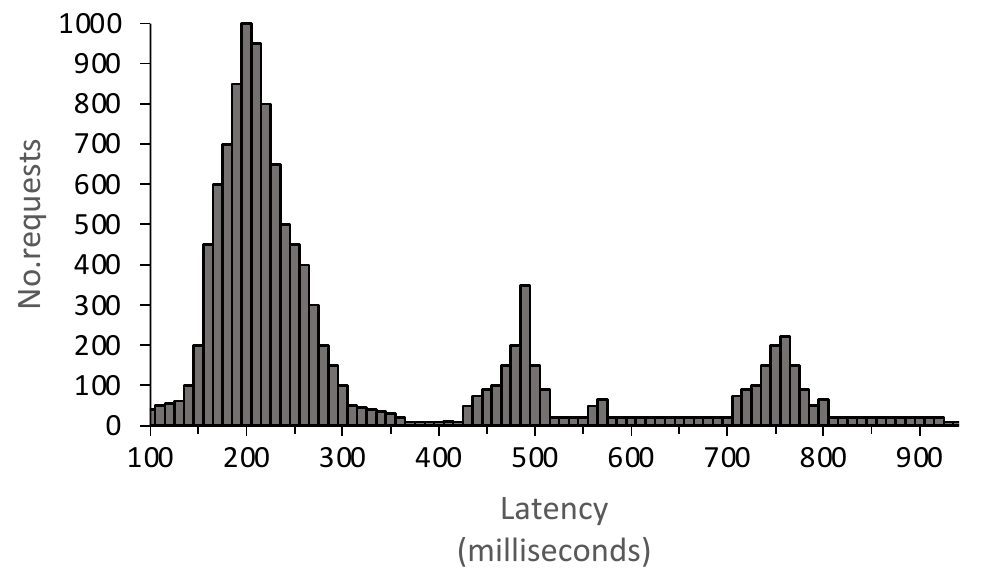}
         \caption{Histogram}
         \label{fig:hist}
     \end{subfigure}
     \caption{Visual Components}
     \label{fig:viscomponents}
\end{figure}

\subsubsection{Tree}
This visualization component takes inspiration from the Jaeger comparison tool~\cite{farro2018}, which allows users to compare two end-to-end requests and highlight their structural differences. We have redesigned this approach by extending its capabilities beyond the comparison of two requests, thereby allowing aggregated analysis of multiple end-to-end  requests.
In a nutshell, the \toolname tree provides an aggregated view of the RPC workflows performed by a set of end-to-end requests, as shown in \figref{fig:tree}.
Each node of the tree represents a RPC invocation within a specific execution path, where the leftmost node represents the root RPC, and edges indicate direct RPC invocation.
For instance, in \figref{fig:tree} the node labeled as \texttt{RPC$_E$} represents the execution path \texttt{RPC$_A$}$\rightarrow$\texttt{RPC$_B$}$\rightarrow$\texttt{RPC$_E$}.
As can be observed by the figure, the same RPC can appear in multiple nodes (\eg \texttt{RPC$_F$}), as it can be invoked within multiple different execution paths.
A RPC execution path will appear in the tree if and only if it is present in at least one of the requests being analyzed.
It is worth noting that when a particular RPC invokes the same RPC multiple times, this leads to a single node in the tree. In other words, if the \texttt{RPC$_A$} invokes the \texttt{RPC$_D$} multiple times, there will be only one child node referring to \texttt{RPC$_D$}.

\toolname utilizes color encoding to highlight RPC execution paths that are worthy of investigation based on their attribute values.
It currently supports the analysis of two kinds of attributes: \emph{execution time} and \emph{frequency}.
The first one denotes the (average) execution time of the RPC within a specific execution path in each request, while second one indicates the path frequency, \ie how many times it occurs within each request.
We use color encoding to emphasize RPC execution paths with higher variance in their attributes.
The key intuition here is that RPC execution paths showing higher variance in their attributes are likely to manifest different behaviors that can potentially affect the end-to-end response time.
For instance, a higher frequency of a particular RPC invocation within a request could result in a longer end-to-end response time. Or similarly, a slower RPC execution time may correspond to a prolonged end-to-end response time. 

We employ a continuous color scale to depict the variability in the attribute values associated.
This scale is based on the Coefficient of Variation (CV)\cite{everitt1998}, \ie a standardized measure of dispersion that is defined as the ratio of the standard deviation to the mean.
As execution times in distributed systems are well known to be subject to long tails~\cite{Dean2013}, when dealing with this attribute, we apply outlier filtering by removing execution times values greater than the 99$^{th}$ percentile.
A CV of 0 results in a white node, indicating no variability. On the other hand, a CV greater than or equal to 1 results in a red node, suggesting a high variability in the attribute values.
The shade of color gradually transitions from white to red as the CV value increases.

\subsubsection{Histogram}

The \toolname histogram component (shown in \figref{fig:hist}) depicts a traditional distribution plot of the end-to-end response time.
These kinds of visualizations are frequently used in practice for performance analysis, and are provided by several tools, \eg Kibana~\cite{kibana}.
According to recent research~\cite{Davidson2023}, understanding the distribution of end-to-end response times stands as core activity in modern performance analysis practice.
The histogram component provided by \toolname aims to facilitate this process by supporting the identification of specific performance behaviors that are worthy of investigation.
The user can visually identify ``modes'' in the response time distribution, which indicate meaningful recurring performance behaviors, to start a targeted investigation on these requests, as we will detail in the subsequent subsection.

%% file: interaction.tex
\subsection{Interaction modalities}

\begin{figure}
    \centering
     \begin{subfigure}[b]{0.9\linewidth}
         \centering
         \includegraphics[width=\textwidth]{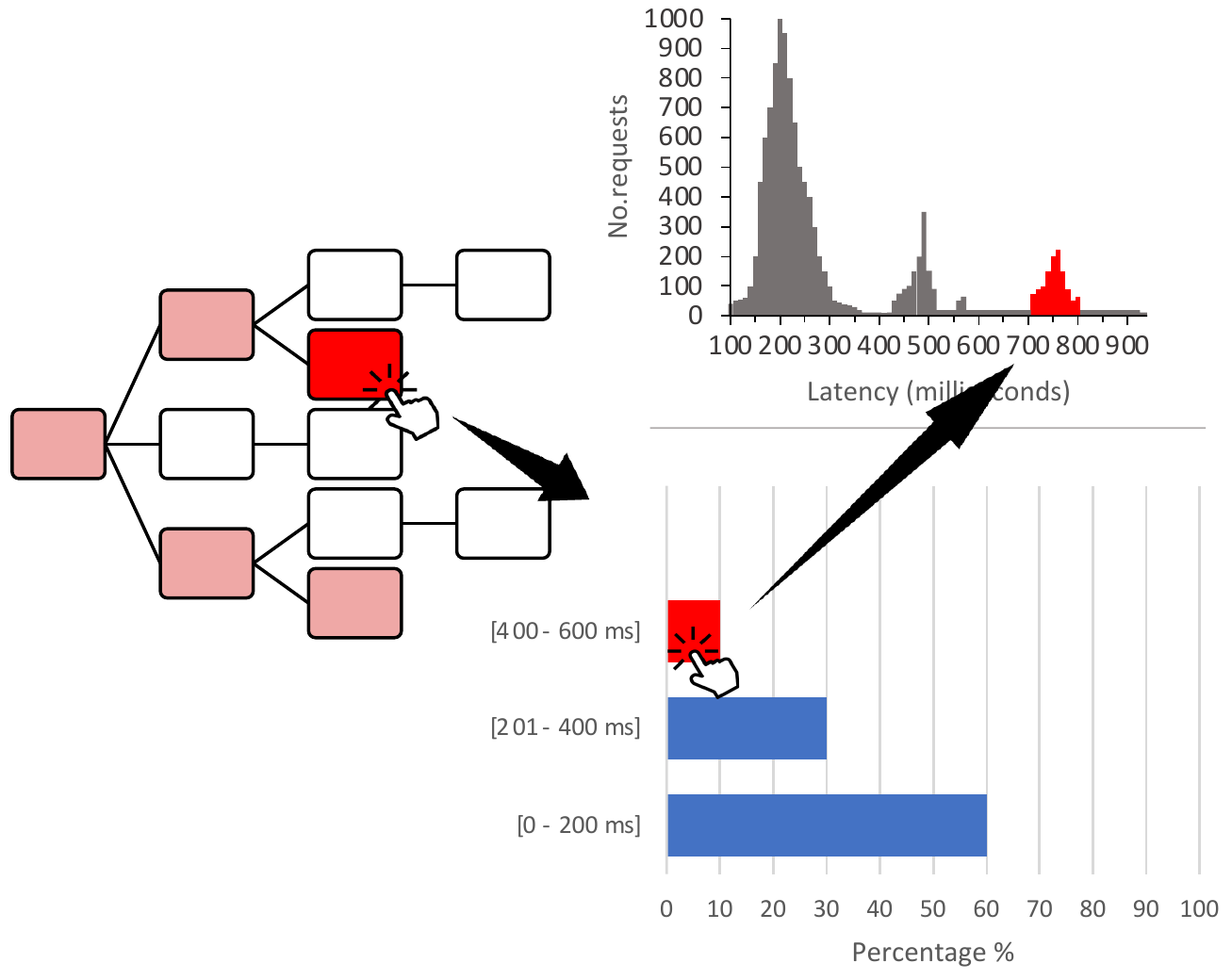}
         \caption{Forward analysis}
         \label{fig:forward}
     \end{subfigure}
     \hfill
     \begin{subfigure}[b]{0.9\linewidth}
         \centering
         \includegraphics[width=\textwidth]{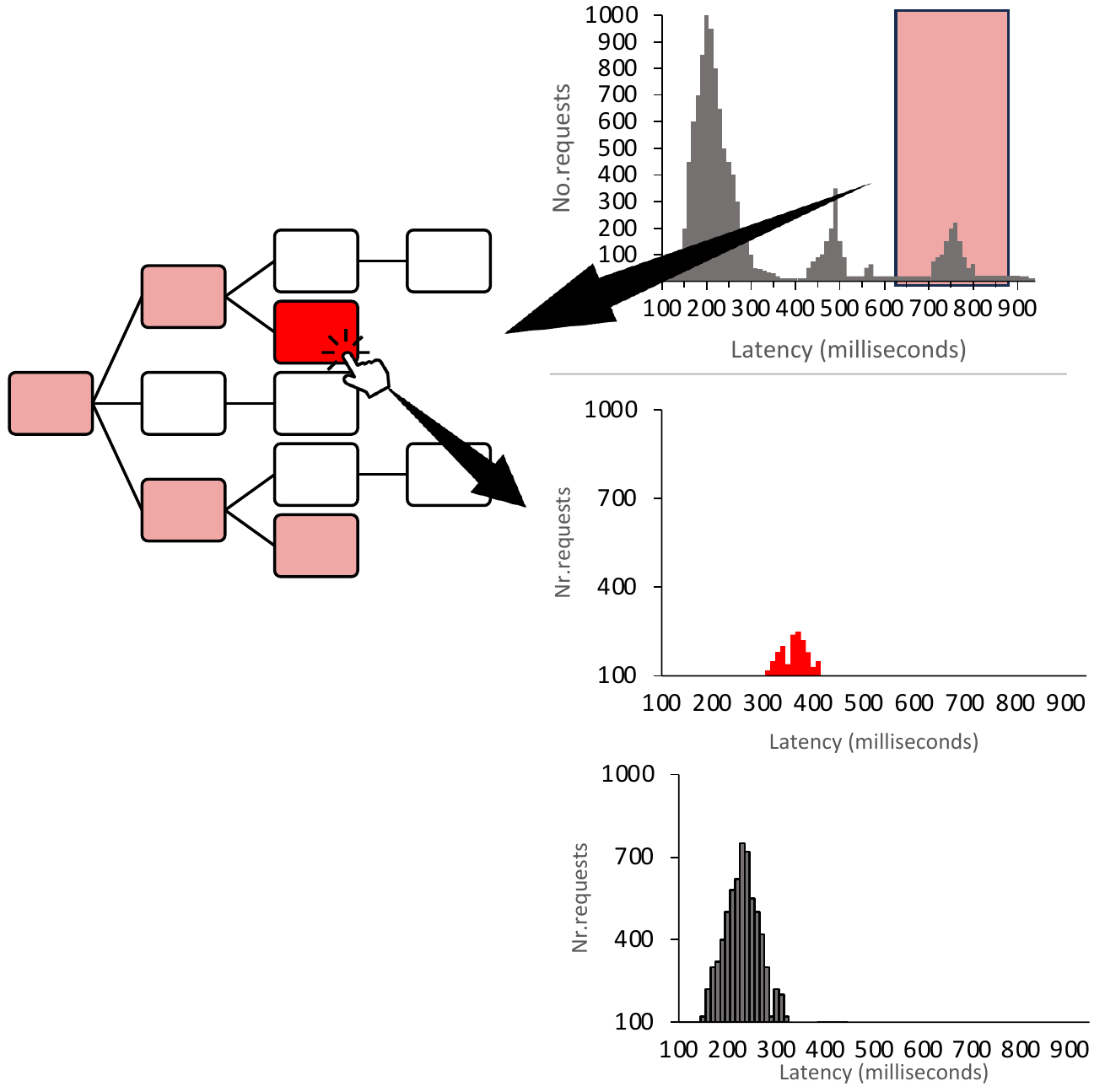}
         \caption{Backward analysis}
         \label{fig:backward}
     \end{subfigure}
     \caption{Interaction modalities}
     \label{fig:interaction}
\end{figure}

\toolname supports bidirectional analysis, allowing users to initiate their analysis from either the tree (\emph{forward analysis}) or the histogram  (\emph{backward analysis}).
In the following, we provide detailed descriptions of both these interaction modalities.

\subsubsection{Forward analysis}
\figref{fig:forward} depicts an illustrative example of forward analysis.
By examining the tree, the user can identify ``suspicious'' RPC execution paths that exhibit high variability in the corresponding attribute values.
For instance, when analyzing the execution time attributes, the user can identify RPCs that show highly varying execution times, and, by clicking on the corresponding node, they can inspect the recurring execution time behaviors associated with the path, displayed in the form of a bar chart, as shown in \figref{fig:forward}.
Each bar refers to a specific execution time range (see y-axis labels), and it shows the percentage of requests with RPC execution time falling in that range.
In order to identify meaningful ranges, we employ a widely-used clustering algorithm, namely K-means~\cite{Lloyd1982}.
In particular, we run the algorithm on-the-fly after the user click with $k$ ranging from 2 to 5 and we select the results showing the highest silhouette score \cite{Rousseeuw1987}.
Each bar represents a meaningful recurring execution time behavior, and the user can click on each bar to see how this behavior reflects in the end-to-end response time.
This relation is shown by highlighting in red the area of the distribution that shows this particular RPC execution time behavior.
For instance, in \figref{fig:forward}, we can observe that when the selected RPC has an execution time ranging between 400 and 600 milliseconds, it can lead to end-to-end response times that range between 700 and 800 milliseconds.
Understanding these kinds of relationships would have been way more challenging by using currently available tools.
It is worth noticing that the same interaction modality also applies when analyzing different RPC attributes, such as \emph{occurences}.

\subsubsection{Backward analysis}
In the backward analysis, the user can start its investigation directly from the histogram component.
The user selects a specific range of end-to-end response time using a slider selector, as shown in \figref{fig:backward}.
This selection triggers an update in the tree component's color scheme, shifting its semantic from variability to divergence.
In other words, the updated color scheme will now denote the degree of divergence in the attribute values of the selected set of requests (\ie those that show end-to-end response time in the selected range) when compared to those in other requests.
A red node indicates that the corresponding RPC execution path shows considerably different attribute values in the selected requests when compared to other requests, suggesting a possible relationship between the selected end-to-end response time and the RPC execution path.
Conversely, a white node indicates similar attribute values, and therefore a weak relation.
We quantify the degree of divergence using Kullback-Leibler divergence~\cite{Boyd2004}, where values close to 0 indicate nearly identical distributions (white), while values close to or higher than 1 indicate highly different distributions (red).

The user can then delve deeper into each RPC execution time behavior by clicking on the corresponding node.
This action lets appear at screen two new histograms (in the bottom right corner) representing the distributions of the execution time in the selected RPC execution path, respectively in the selected requests (in red) and in other requests (in grey).
In doing so, the user can effectively analyze how particular ranges of the end-to-end response time distribution correlate with specific RPC attribute values.

%% file: dashboard.tex
\subsection{Dashboard}
\begin{figure}
    \centering
    \includegraphics[width=\linewidth]{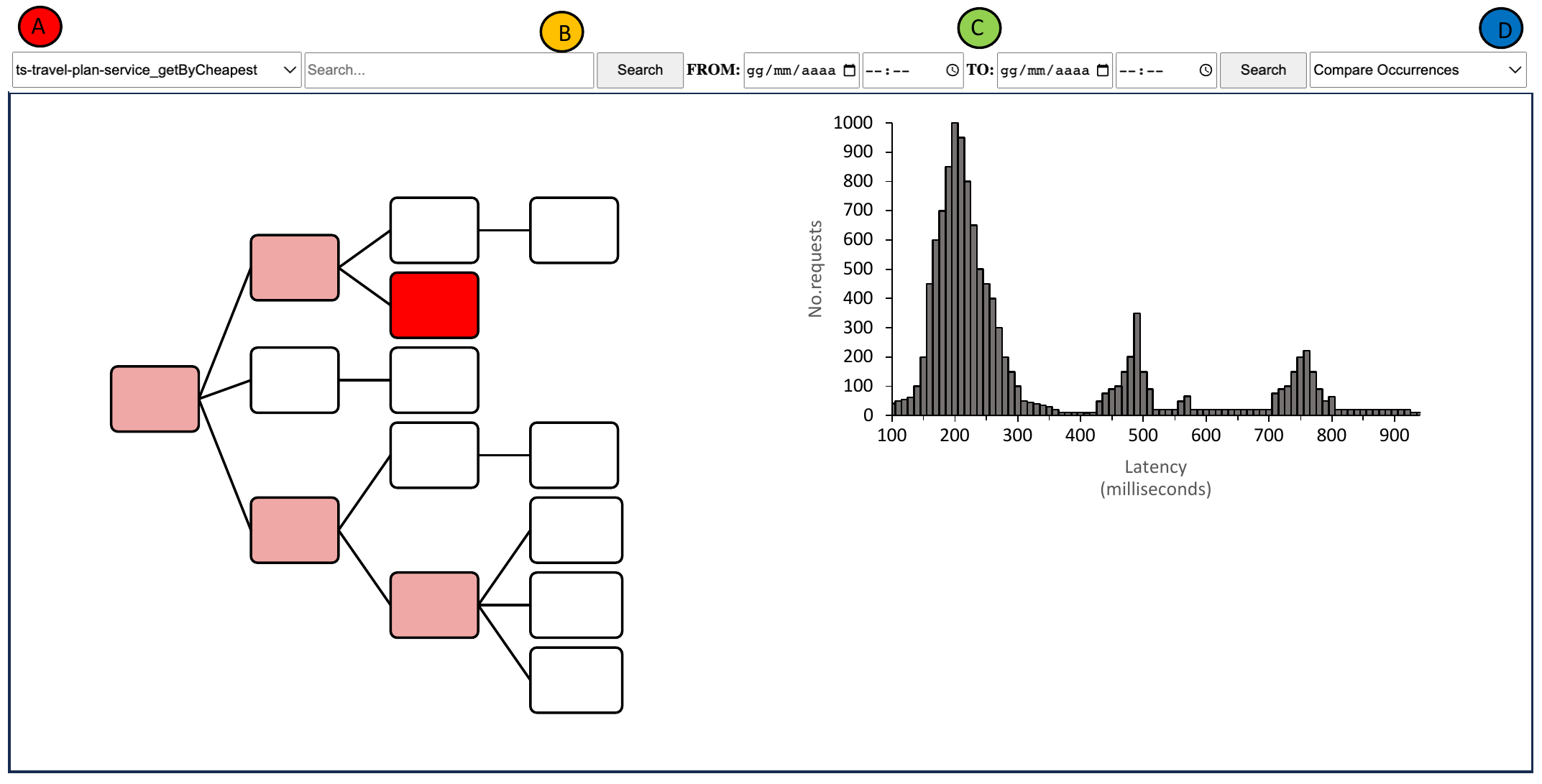}
    \caption{\toolname's Dashboard}
    \label{fig:dashboard}
\end{figure}

\figref{fig:dashboard} outlines the \toolname dashboard. As can be observed by the figure the two main visual components, namely the tree and the histogram, are positioned in the center-left and in the upper-right corners, respectively. The space in the bottom-right is intentionally left blank and will be used to display supplementary visualization components during the interaction, \eg the bar chart (for forward analysis) and the two histograms (for backward analysis).

It's worth noting that \toolname is specifically designed to assist in analyzing requests from the same class, \ie those originating from the same root RPC. As part of this process, the user  is required to first select the root RPC and the RPC attribute (\ie execution time or path frequency) to be investigated, before proceeding with the actual analysis.

The user can select the root RPC using either a dropdown menu \ccircle{red}{A} or a search text-box \ccircle{yellow}{B}.
Similarly, the RPC attribute (execution time or frequency) to be analyzed can be selected using a dropdown menu \ccircle{blue}{D}.
Additionally, the dashboard includes a date-time range selector \ccircle{green}{C}, where the user can specify the start and end date-times.
This feature allows for analyses at different time granularities (\eg monthly, weekly, and daily) or over specific time ranges known to include system anomalies.

To enhance user experience during the interaction with the tool, \toolname supports pinch gestures to enable zoom in and zoom out of the tree.
In addition, it allows the user to hide the RPCs invoked within a particular execution path by double-clicking on the related node.

%% file: implementation.tex
%%%%%%%%%%%%%%%%%%%%%%%%%%
%%%%%%%%%%%%%%%%%%%%%%%%%%
\subsection{Architecture and Implementation}
%%%%%%%%%%%%%%%%%%%%%%%%%%
%%%%%%%%%%%%%%%%%%%%%%%%%%

\begin{figure}
    \centering
    \includegraphics[width=\linewidth]{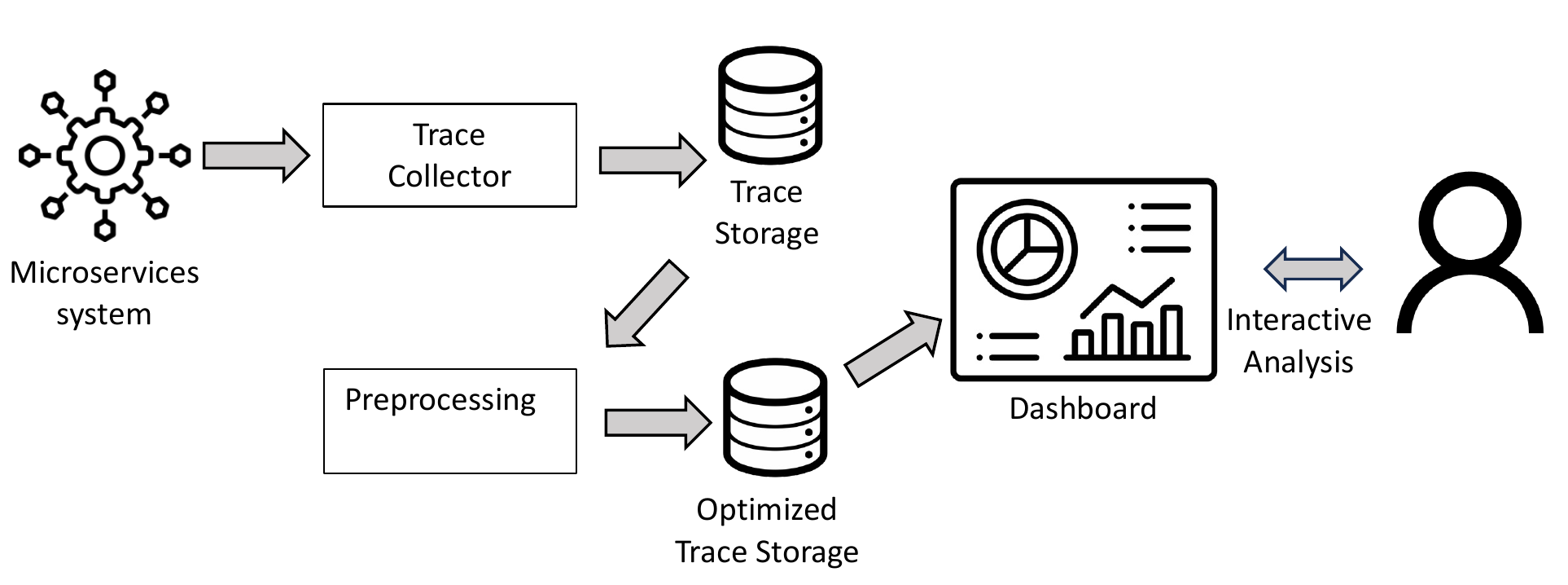}
    \caption{\toolname's Workflow}
    \label{fig:workflow}
\end{figure}

\figref{fig:workflow} outlines the key architecture components of \toolname.
The \emph{Trace Collector} (\ie Jaeger~\cite{jaeger}) continuously collects traces from the microservices system and stores them in a \emph{Trace Storage} (\ie Elasticsearch~\cite{elasticsearch}).
Given the large volume of data collected each day, we have devised a \emph{Preprocessing} step to enhance the efficiency of interaction with \toolname.
This preprocessing step operates in batches and is intended to be executed periodically (\eg hourly or daily).
For each end-to-end request (\ie trace), \toolname recursively reconstructs all the involved RPC execution paths, along with their attribute values (namely, execution times and frequencies), and stores them in an \emph{Optimized Trace Storage} based on MongoDB.
Each path associated with a request is stored as a separate document in a dedicated MongoDB collection, and includes: the name of the path, the trace ID, the number of occurrences of the path in the trace, the observed execution time, the timestamp, and the name of the root RPC.
Similarly, \toolname stores the end-to-end response time values, along with related information, in a separate MongoDB collection. This information includes the RPC root, the trace ID, the response time value, and the timestamp.
This data reorganization allows for greater flexibility in easily and efficiently querying the data needed for the \toolname dashboard to function properly.
As can be seen from \figref{fig:workflow}, the \emph{Dashboard} app directly queries the \emph{Optimized Trace Storage} to efficiently generate visualizations.

\toolname currently supports distributed traces stored in the Jaeger \cite{jaeger} format using Elasticsearch \cite{elasticsearch} as \emph{Trace Storage}, but it can be easily extended to other technologies. The dashboard and visual components have been developed using D3.js, which handles the visualization rendering, and Flask, which serves as the backend service.
The preprocessing scripts are implemented in Python.

%% file: evaluation.tex
%%%%%%%%%%%%%%%%%%%%%
%%%%%%%%%%%%%%%%%%%%%
\section{Evaluation}
%%%%%%%%%%%%%%%%%%%%%
%%%%%%%%%%%%%%%%%%%%%
The conducted evaluation is centered around one main research question:
\emph{To what extent does \toolname support performance analysis? } We want to understand whether \toolname can be successfully utilized to gain insights about the relationship between request attributes and end-to-end performance response time. 

In the following, we first describe the methodology used to gather the answer. Then, we report and discuss the results of the experimental evaluation. Finally, we describe the threats to validity of our study.

\input{methodology}

\input{results}

\input{threats}

\input{replication}

%% file: methodology.tex
%%%%%%%%%%%%%%%%%%%%%
\subsection{Methodology}
%%%%%%%%%%%%%%%%%%%%%

%Our \emph{goal} is to evaluate to what extent \toolname can assist in understanding the relationship between request attributes and end-to-end response time.
To achieve our study \emph{goal}, we generate \datasets datasets of distributed traces, where each dataset reflects a distinct scenario that induces a specific variation in the relationships between request attributes and end-to-end response time. Subsequently, we manually analyze each dataset using \toolname to evaluate the effectiveness of our tool in highlighting these relationships.

\subsubsection{Datasets generation}
The \datasets datasets are generated from TrainTicket \cite{zhou2021}, which, as best as we know at the time of writing, is the largest and most complex open-source microservice-based system. TrainTicket provides a typical train ticket booking web service; it involves 41 microservices implemented in four programming languages, and it utilizes Jaeger~\cite{jaeger} and Elasticsearch~\cite{elasticsearch} for collecting and storing distributed traces.
We have chosen TrainTicket as a representative case system due to its complexity and because it has been recently used in software engineering research~\cite{traini2022b, zhang2022,li2022, zhou2021}.

Each dataset of our study contains distributed traces related to one specific \emph{root RPC} of the system, which are stored on Elasticsearch using the standard Jaeger format.

To simulate different scenarios that induce different variations in the relationship between RPC attributes and end-to-end response time, we rely on two different approaches: (i) we inject synthetic performance issues in specific RPCs to increase the overall end-to-end response time, and (ii) we use complex mixtures of varying workloads that may alter the relationships between RPC execution time/occurrences and end-to-end response time.
These two distinct approaches lead to the generation of two categories of datasets.

The first category of datasets is generated using a methodology similar to the one presented in \cite{traini2022b, cortellessa2020}. Initially, the system's source code is modified to inject random performance issues. Following this, load-testing sessions are run to simulate user interactions with the system and generate distributed traces. Each injected performance issue affects approximately 10\% of requests, introducing a delay into one specific RPC.

To generate a dataset, we first select two random RPCs that will be impacted by the performance issues. Subsequently, we choose a random delay to increase the end-to-end response time by $x\%$, where $x\in\{10, 20, 30\}$.
In addition, in half of the datasets,  we inject a random delay of $y\%$ (with $y\in\{10, 20, 30\}$) into an asynchronous RPC, which does not produce any effect on the end-to-end response time.
This is a common practice used to test the robustness of pattern detection approaches in the context of microservices systems~\cite{cortellessa2020,traini2022b, krushevskaja2013}.
After modifying the system accordingly, we conduct load-testing sessions to generate the distributed trace datasets.
Each load testing session involves 20 synthetic users, simulated by Locust~\cite{locust}. Each user makes a request to the system and randomly waits between 1 and 3 seconds before making the next request. Each session lasts for 20 minutes.
Using this methodology, we generate 20 datasets featuring various combinations of performance issues that affect different RPCs with different delays. For a more detailed explanation of this process, we refer readers to the work of Traini and Cortellessa~\cite{traini2022b}. Due to space constraints, we do not elaborate further here.

The second kind of dataset does not involve any performance issue injection, but it is generated using a more elaborate workload generator.
Similarly to recent studies~\cite{liao2021,liao2020} we use load mixtures that involve multiple types of simulated users (\ie load drivers), where each user type performs different classes of requests on the system. For example, some types of users may only visit the homepage and subsequently search trains for some random locations, while others first login into the system and then book random tickets. Besides this, we also ensure that the number of simulated users per type keeps changing over time. In this way, workloads will more closely resemble real-world ones, as they generate mixtures of different classes of requests that change over time \cite{ardelean2018}.
To this aim, we slightly modified \textit{PPTAM}~\cite{avritzer2019}, a workload generator that involves 5 different user types, to continuously change the number of users of each type at run-time.
Overall, the number of simultaneous users ranges from a minimum of 20 to a maximum of 31, and the load-testing session lasts for 1 hour.
The workload fluctuations over time are randomly generated upfront.
This process leads to 13 distinct datasets, each one related to a different API.

The generations of the datasets were done on a bare-metal machine running Linux Ubuntu 18.04.2 LTS on a dual Intel Xeon CPU E5-2650 v3 at 2.30 GHz, with a total of 40 cores and 80 GB of RAM. All non-mandatory background processes except SSH are disabled, and we ensured that no other users interacted with the dedicated machine during our experiments.

To enhance clarity throughout the rest of the article, we will use specific notations for different categories of datasets. Datasets characterized by performance issues (\ie first category) will be referred to as $\widehat{D}_i$, where $1 \leq i \leq 20$. Conversely, datasets free from performance issues (\ie second category)  will be denoted as $D_i$, with $1 \leq i \leq 13$.

\subsubsection{Manual analysis}
To assess the effectiveness of our approach, two authors conducted manual inspections of the \datasets  distributed trace datasets using \toolname.
Our evaluation focused on determining the extent to which \toolname facilitated the comprehension of the relationship between RPC execution time/frequency and end-to-end response time.

It is worth noting that neither author was aware of the specific performance issues or the workload variations present in each dataset. This is because the process for the dataset generation, including performance issue injections and load testing modifications, was entirely random and automated.
Nonetheless, both authors were familiar with the TrainTicket system before the study.

%For the first type of dataset, our evaluation focused on determining the extent to which \toolname facilitated the identification of the injected performance issues.
%Conversely, for the second type of dataset, our evaluation centered around assessing \toolname's ability to support the understanding of specific end-to-end response time behaviors.

%% file: results.tex
%%%%%%%%%%%%%%%%%%%%%
\subsection{Results}
%%%%%%%%%%%%%%%%%%%%%
%\LUCA{Work in progress}

\toolname has proven to be effective in highlighting the relationship between RPC execution time and end-to-end response time, throughout all the datasets featuring injected performance issues.
The analysis was straightforward for the majority of the datasets (18 out of 20), demanding minimal interaction with \toolname.
In these datasets, both \emph{forward} and \emph{backward analysis} demonstrated comparable effectiveness, with no noticeable difference in the effort needed to understand these relationships.
Due to space constraints, we are unable to present the exhaustive results of our analyses across all datasets.
However, we have included a selection of representative examples that underscore both the utility and potential challenges associated with employing \toolname.
Additionally, for the sake of completeness, we have made available screenshots capturing interactions with \toolname across all the datasets in a supplementary replication package \cite{replication}.

\begin{figure*}
    \centering
    \includegraphics[width=\linewidth]{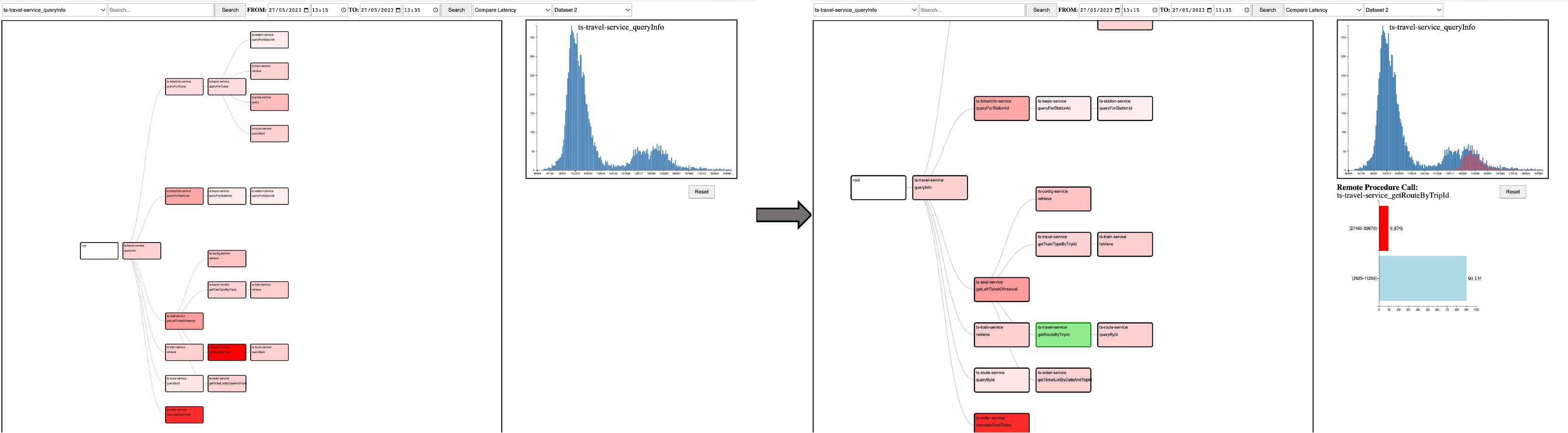}
    \caption{\emph{Forward analysis} on \emph{execution time} for dataset $\widehat{D}_2$.}
    \label{fig:forward_result}
\end{figure*}

\figref{fig:forward_result} showcases an example of \emph{forward analysis} using the dataset $\widehat{D}_2$. As depicted in the left screenshot, \toolname significantly streamlines the identification of the two RPCs impacted by performance issues, \ie the ones highlighted in bright red.
Following this, the user can select these nodes to investigate correlations between specific RPC execution times and end-to-end response times.
For example, the screenshot on the right of \figref{fig:forward_result} reveals that the selected RPC execution path (highlighted in green) exhibits two distinct execution time behaviors: in 9.87\% of the requests, the RPC \rpc{getRouteByTripId} has an execution time ranging from 27.46 to 33.67 milliseconds,  and in the remaining 90.13\% of requests, the execution time ranges from 2.62 to 11.25 milliseconds.
This screenshot displays the view of \toolname during the investigation of the first behavior, that is, after clicking on the corresponding bar (highlighted in red). As evident from the figure, \toolname reveals that all the requests with an execution time ranging from 27.46 to 33.67 milliseconds in the selected RPC execution path fall within a specific region of the end-to-end response time distribution, as shown by the red highlight in the histogram.
Understanding these kinds of relationships would have been particularly challenging when using traditional performance analysis tools.

\begin{figure*}
    \centering
    \includegraphics[width=\linewidth]{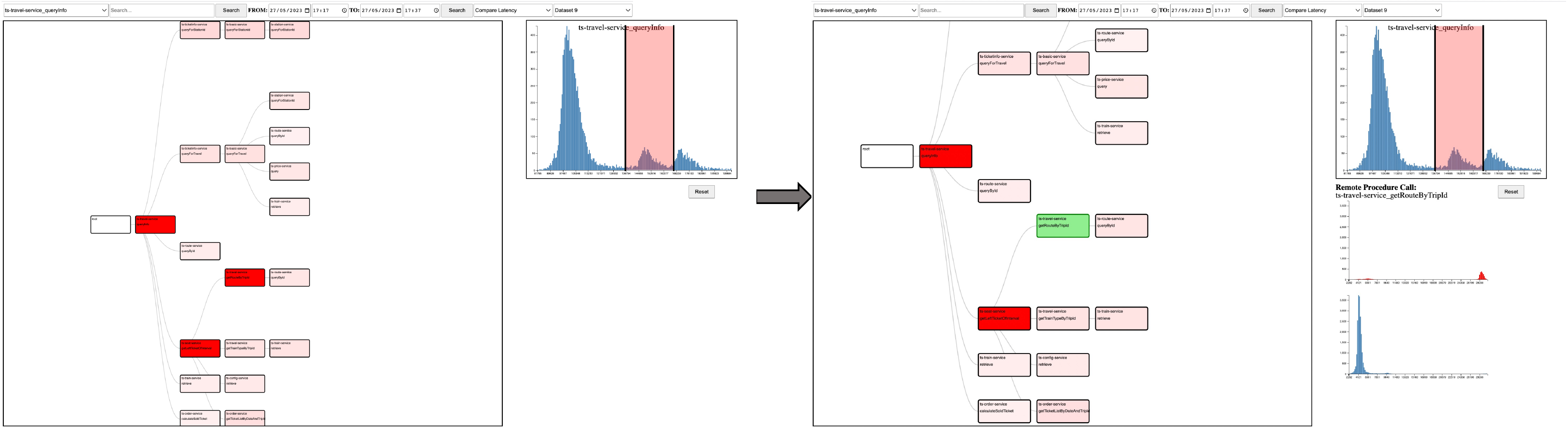}
    \caption{\emph{Backward analysis} on \emph{execution time} for dataset $\widehat{D}_9$.}
    \label{fig:backward_result}
\end{figure*}

\figref{fig:backward_result} offers another example of how \toolname allows users to rapidly identify the RPC responsible for a particular end-to-end response time deviation.
Specifically, this figure demonstrates an instance of a \toolname \emph{backward analysis} using the dataset $\widehat{D}_9$.
The left screenshot shows that by selecting a specific range of end-to-end response times, the user can immediately pinpoint the RPC execution paths that display significantly divergent behavior in the execution time (highlighted in bright red).
The screenshot on the right displays the investigation of one of these nodes (\ie the one highlighted in green),  illustrating how \toolname assists users in comprehending the correlation between specific RPC execution times and the selected range of end-to-end response times.
For instance, it is noticeable that when the RPC \rpc{getRouteByTripId} has an execution time exceeding 27 milliseconds, it results in an end-to-end response time that falls within the range of 137 and 168 milliseconds.

\begin{figure}
    \centering
     \begin{subfigure}[b]{0.98\linewidth}
         \centering
         \includegraphics[width=\textwidth]{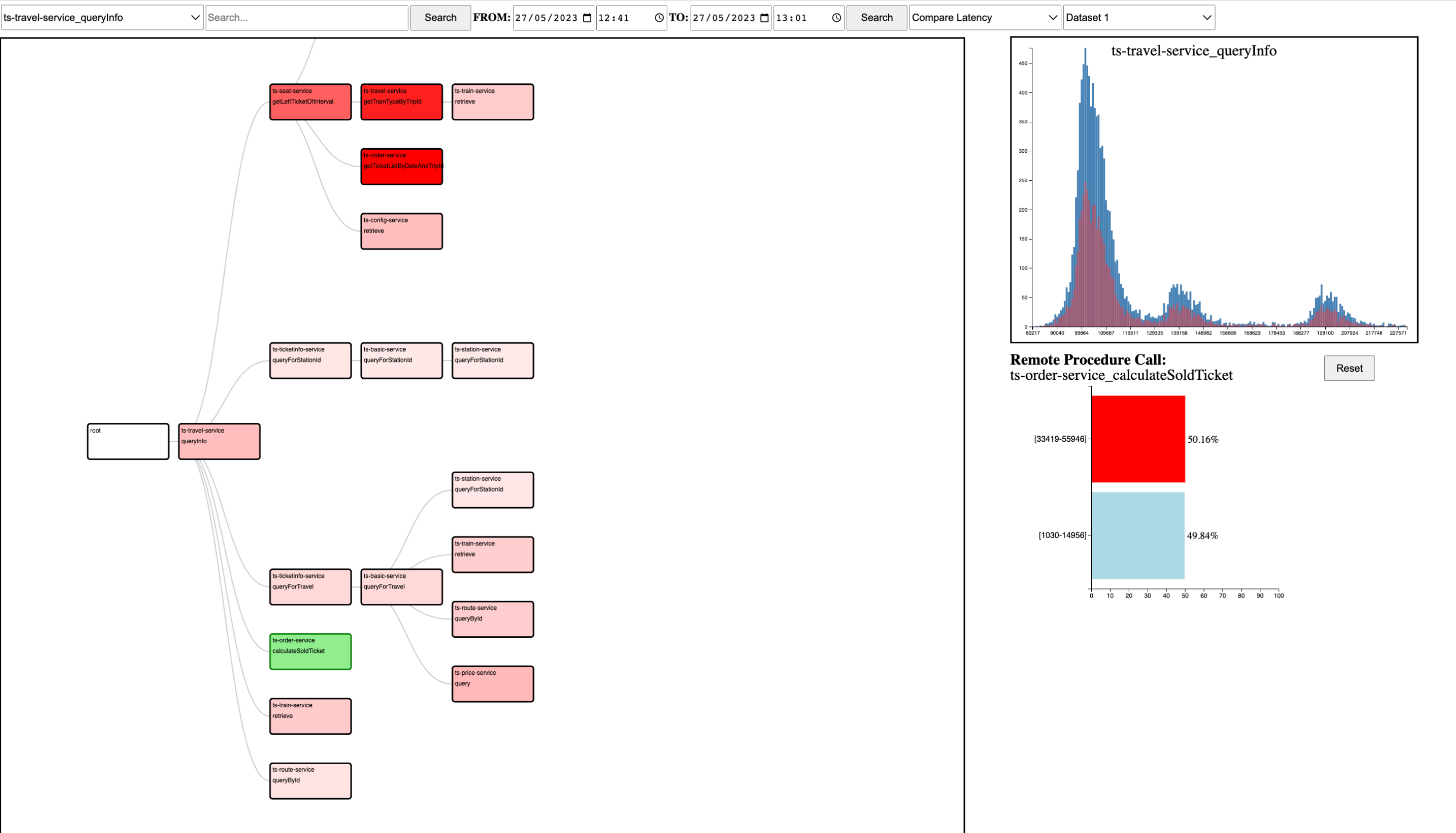}
         \caption{ }
         \label{fig:non_critical_rpc_1}
     \end{subfigure}
     \hfill
     \begin{subfigure}[b]{0.98\linewidth}
         \centering
         \includegraphics[width=\textwidth]{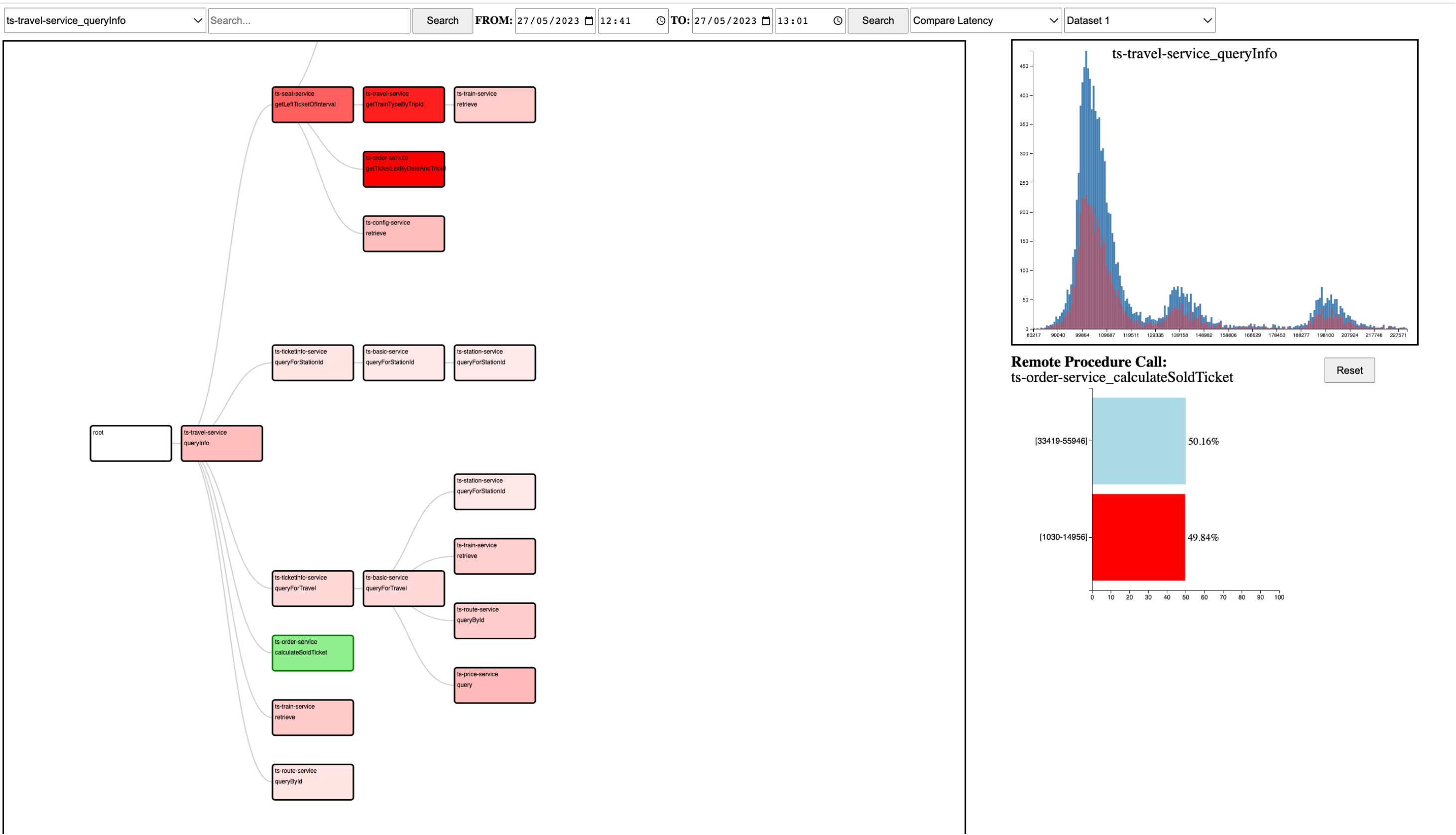}
         \caption{ }
         \label{fig:non_critical_rpc_2}
     \end{subfigure}
     \caption{\emph{Forward analysis} on \emph{execution time} for dataset $\widehat{D}_1$.}
     \label{fig:non_critical_rpc}
\end{figure}

Another interesting aspect of \toolname is its ability to identify execution time fluctuations in RPCs that do not have influence on the end-to-end response time.
For instance, \figref{fig:non_critical_rpc} illustrates two distinct execution time behaviors in the selected RPC \rpc{calculateSoldTicket}: one ranging between 33.42 and 55.95 milliseconds, and another between 1.03 and 14.96 milliseconds.
Through the use of \toolname, we were able to easily notice the lack of correlation between the execution time of this RPC and the end-to-end response time.
As illustrated in Figures \ref{fig:non_critical_rpc_1} and \ref{fig:non_critical_rpc_2}, the selected execution time behaviors (\ie the bars highlighted red) are evenly distributed across the end-to-end response time distribution, implying a lack of notable correlation with specific regions of the end-to-end response time.
This indicates that even if the RPC execution time varies drastically from one request to another, it does not have any significant impact on the end-to-end response time.

In two datasets, specifically $\widehat{D}_4$ and $\widehat{D}_{19}$, the analysis was more complex, requiring a higher number of interactions with \toolname.
The peculiarity of these datasets was that the two performance issues led to an increased  end-to-end response time that overlaps within the same range. This made the connection between the RPC execution time and end-to-end response time more challenging to understand.
 We did not include the specific details of these cases in the paper because of limited space, but we refer the reader to our supplementary materials \cite{replication} for the related screenshots.

\begin{figure}
    \centering
     \begin{subfigure}[b]{0.98\linewidth}
         \centering
         \includegraphics[width=\textwidth]{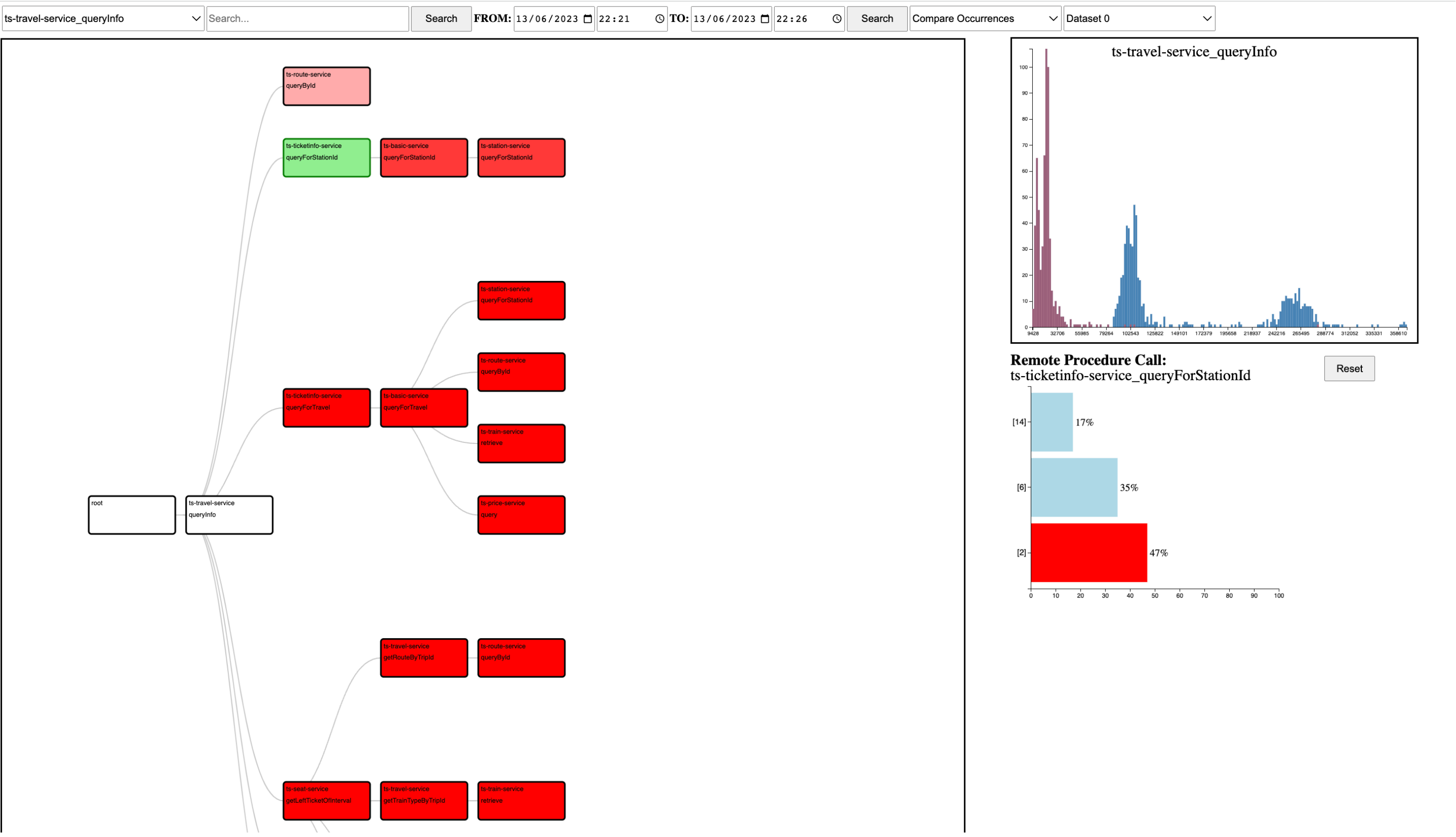}
         \caption{}
         \label{fig:occ_analysis_1}
     \end{subfigure}
     \hfill
     \begin{subfigure}[b]{0.98\linewidth}
         \centering
         \includegraphics[width=\textwidth]{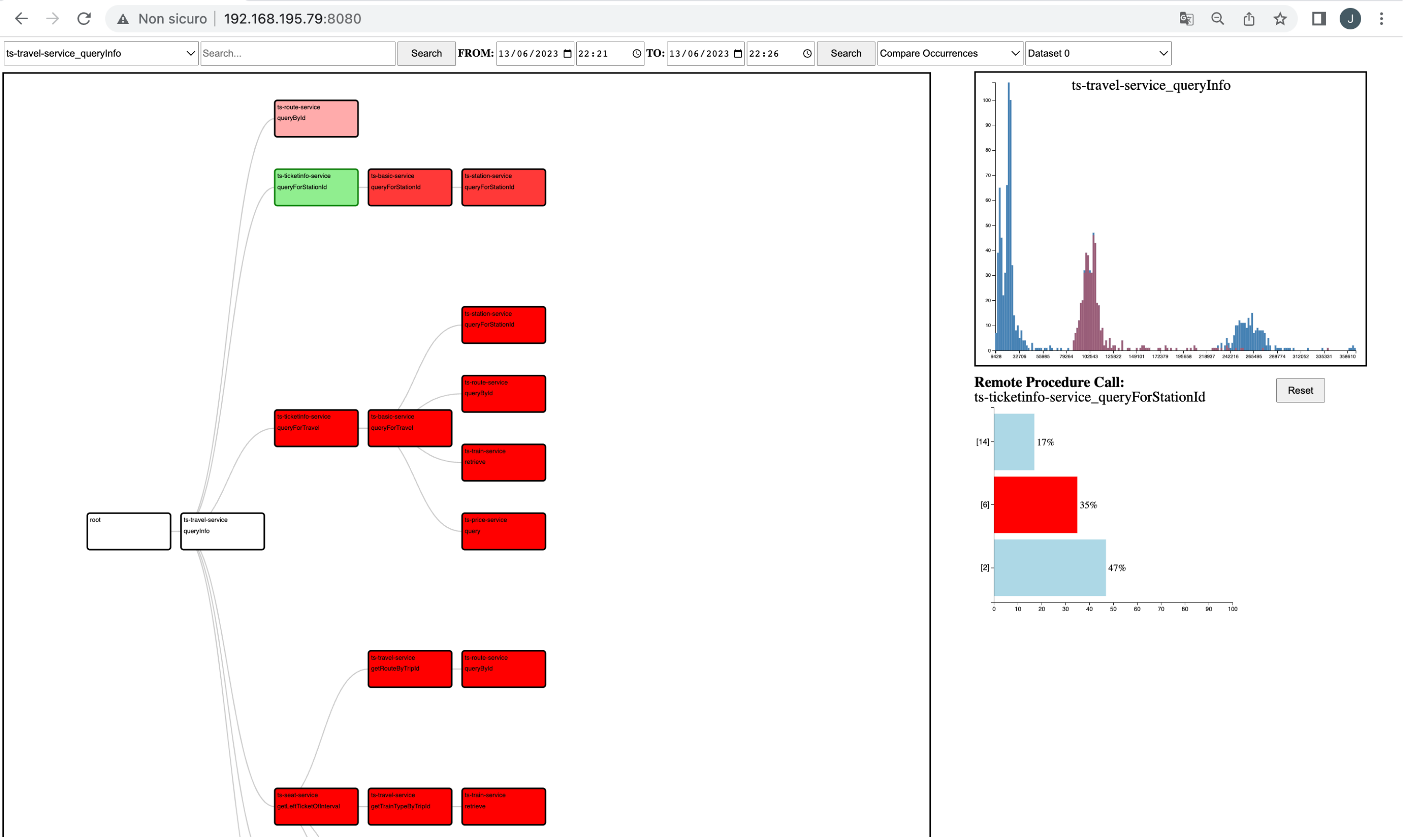}
         \caption{}
         \label{fig:occ_analysis_2}
     \end{subfigure}
     \hfill
     \begin{subfigure}[b]{0.98\linewidth}
         \centering
         \includegraphics[width=\textwidth]{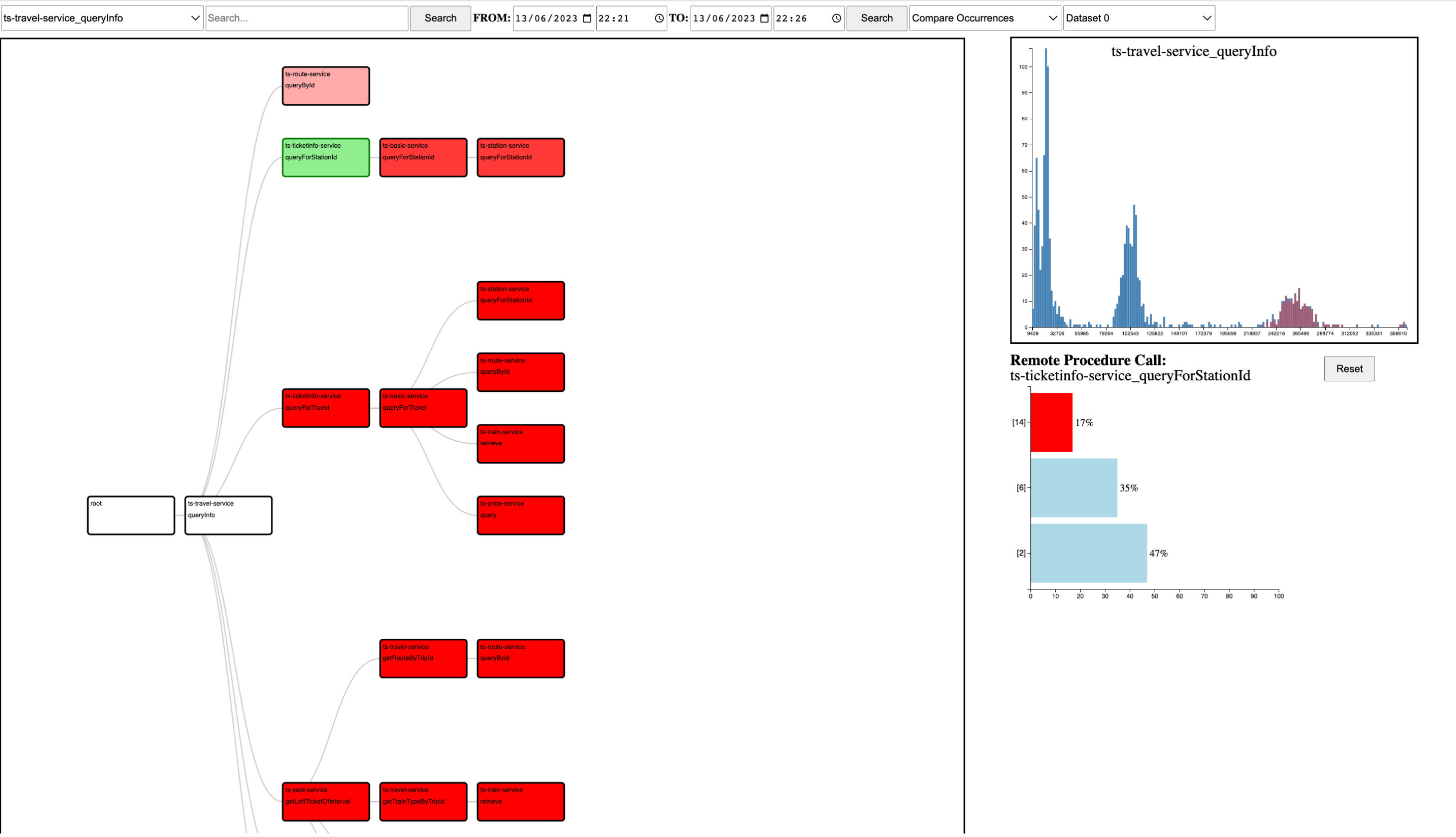}
         \caption{}
         \label{fig:occ_analysis_3}
     \end{subfigure}
     \caption{\emph{Forward analysis} on \emph{frequency} for dataset $D_{2}$.}
     \label{fig:occ_analysis}
\end{figure}

With regards to the 13 datasets in the second category, we found that a substantial majority of them - precisely 11 datasets - feature a unique mode in the end-to-end response time distribution. Given the objectives of our analysis, these cases were not considered. Consequently, we used two datasets for our evaluation. The first dataset, $D_1$, consists of requests originating from the root RPC \rpc{getByCheapest}, while the second dataset, $D_2$, comprises requests initiated from \rpc{queryInfo}.
\toolname enabled us to characterize the correlation between the frequency of each RPC execution path and specific modes of the end-to-end response time.
\figref{fig:occ_analysis} provides an example of this characterization, illustrating that each mode of the end-to-end distribution corresponds to a specific number of invocations of a selected RPC execution path (highlighted in green).
For example, as depicted in \figref{fig:occ_analysis_3}, the right-most mode is characterized by 14 invocations of the path \rpc{queryInfo} $\rightarrow$ \rpc{queryForStationId}. Similarly, the center mode is characterized by 6 invocations of this path, while the left-most mode is marked by 2 invocations.
Uncovering such patterns using traditional observability tools would have been notably challenging.

Summing up, we answer our RQ as follows: \toolname proved to be effective in supporting performance analysis of microservices.
In 18 out of the 20 datasets involving performance issues, we were able to rapidly identify the affected RPCs, their corresponding execution time behaviors, and their relationship with end-to-end response time. However, in a few specific cases (2 out of 20 datasets), the analysis proved to be more challenging, necessitating a greater number of interactions with \toolname.
Moreover, our evaluation demonstrates how \toolname can facilitate an understanding of how structural differences in requests (\ie varying frequencies of RPC execution paths) influence end-to-end response time.
%In both datasets utilized for our evaluation, we successfully disclosed the impact of the frequency of each RPC execution path on end-to-end response time.

%% file: threats.tex
%%%%%%%%%%%%%%%%%%%%%
\subsection{Threats to Validity}
%%%%%%%%%%%%%%%%%%%%%

\subsubsection{Construct validity}

We conducted the evaluation in-house rather than using external participants. Our familiarity with the tool and the experimental setup could potentially introduce bias into the evaluation outcomes. To mitigate this threat, we generated 20 diverse scenarios randomly, each involving various combinations of performance issues. Additionally, the two authors who conducted the evaluations were kept uninformed about the RPCs affected by the performance issues.
Using artificial delays as part of the evaluation may not perfectly mirror real-world performance issues. However, this methodology aligns with prevailing practices in software engineering research, as evidenced by several studies \cite{traini2022b, liao2021, laaber2018, luo2021}. Furthermore, in contrast to many previous studies \cite{liao2021, laaber2018, luo2021}, which typically employ a limited set of predetermined regressions with fixed magnitudes, our approach offers a more comprehensive evaluation on 20 diverse scenarios involving different combinations of RPCs and delay magnitudes

\subsubsection{Internal validity}

The workloads used in our experimental setup may not be representative of real-world workloads. To (partially) mitigate this limitation, we perform an additional analysis using mixtures of continuously changing workloads, generated through PPTAM \cite{avritzer2019}.
Our evaluation may be subject to confirmation bias, wherein the authors may unconsciously confirm their pre-existing beliefs on the effectiveness of \toolname.
Nonetheless, the results obtained using \toolname unambiguously demonstrate its effectiveness across a majority of the datasets evaluated.
In the interest of transparency and to enable readers to independently assess this evidence, we have made all screenshots documenting the use of \toolname across the datasets in our study publicly available~\cite{replication}.

\subsubsection{External validity}
We cannot ensure that \toolname can achieve the same effectiveness on other datasets outside our experimental setup (\eg real world scenarios).
Nevertheless, through an evaluation on \datasets datasets, we have demonstrated that our approach effectively aids in the performance analysis of microservices systems.
\toolname's efficiency was evaluated on datasets of varying sizes, ranging from 11181 to 22348 requests. It’s worth noting that real-world microservices systems may involve a much larger volume of requests. As part of our future work, we plan to enhance the scalability of \toolname by incorporating sampling techniques and optimizing preprocessing procedures.

%% file: replication.tex
%%%%%%%%%%%%%%%%%%%%%%
%\subsection{Replication package}
%%%%%%%%%%%%%%%%%%%%%%
%
%To aid reproducibility we provide the data and source code needed to replicate our findings~\cite{replication}.

%% file: related.tex
%%%%%%%%%%%%%%%%%%%%%%%%%%
%%%%%%%%%%%%%%%%%%%%%%%%%%
\section{Related work}
%%%%%%%%%%%%%%%%%%%%%%%%%%
%%%%%%%%%%%%%%%%%%%%%%%%%%

Previous research on visualization for distributed systems has primarily focused on analyzing individual requests or comparing two requests.
The swimlane visualization, a widely used technique to represent individual end-to-end executions, was originally proposed by Singelman \etal \cite{sigelman2010}. Today, most distributed tracing tools offer this visualization.
TraVista \cite{anand2020} enhances the standard swimlane visualization by augmenting it with information that assists users in contextualizing the performance of the analyzed request in relation to others.
Beschastnikh \etal \cite{beschastnikh2020} introduced a novel visualization tool called ShiViz, which includes an interactive time-space diagram for visualizing individual end-to-end executions of a distributed system.

Sambasivan et al.~\cite{sambasivan2013} studied and compared three visualization approaches (\ie side-by-side view, difference view, and animation) for comparing two request-flow traces.
Jaeger~\cite{jaeger} provides a feature to visually compare the structural characteristics of two requests~\cite{farro2018}.

Several visualizations have also been introduced to analyze the performance behaviors of multiple end-to-end requests in aggregate.
One example is TransVis by Beck \etal \cite{beck2021}, which provides a visualization technique for specifying and analyzing transient performance behaviors in microservice systems.
Other examples of visual techniques for aggregate performance analysis can be found in commercial APM tools~\cite{ahmed2016}, such as Dynatrace, AppDynamics, or Instana. 
For instance, Dynatrace's Service Flow feature~\cite{dynatraceflow} allows to display aggregate workflows of end-to-end requests along with their associated characteristics.

To the best of our knowledge, despite the many existing visualization techniques for microservices performance analysis, there is still a lack of dedicated visualizations to analyze the correlation between requests' attributes and their end-to-end performance behavior, which is the goal of our study.

%In addition to Dynatrace's Service Flow, there have been other studies that propose automated methods to assist in detecting patterns in request characteristics that are associated with anomalous end-to-end performance behavior~\cite{traini2022b, bansal2020, cortellessa2020, krushevskaja2013}. These studies aim to develop techniques and algorithms that can identify patterns or correlations between request attributes and performance issues.

Other related works include the recent Davidson and Mace's survey~\cite{davidson2022}, which underscores the critical role of visualization within systems research, and the qualitative interview study conducted by Davidson \etal~\cite{Davidson2023}, which highlighted the limitations of current distributed tracing tools. Davidson \etal's findings involved several open research challenges that span multiple research areas, including visualization research.

%% file: conclusion.tex
%%%%%%%%%%%%%%%%%%%%%
%%%%%%%%%%%%%%%%%%%%%
\section{Conclusion}
%%%%%%%%%%%%%%%%%%%%%
%%%%%%%%%%%%%%%%%%%%%

In this paper, we presented \toolname, a novel visual analytics  tool for microservices performance analysis.
\toolname overcomes the limitations of current distributed tracing tools by providing a wide set of interactive visualizations that enables effective performance analysis of multiple end-to-end requests.
Through an evaluation of \datasets datasets generated from an established open-source microservices system, we demonstrate how \toolname can be effectively used to understand the relationship between the RPC attributes and end-to-end response time.
For future work, %we aim to broaden the capabilities of \toolname to include additional RPC attributes, such as HTTP headers.
%Furthermore, 
we plan to enhance the efficiency of our tool to facilitate its transition to practice.
As part of this process, we intend to validate our future improvements using real-world distributed traces from large-scale microservices systems, similar to those shared by Alibaba~\cite{luo2021}.
%Finally, we plan to deploy \toolname as a service for the   SoBigData research infrastructure and community\footnote{SoBigData is a research infrastructure that has the goal of enhancing interdisciplinary and innovative research on the multiple aspects of social complexity by combining data and model-driven approaches. SoBigData emphasizes the concept of \textit{responsible data science}. Consequently, SoBigData RI develops methodologies and approaches to put into practice the FAIR (Findable, Accessible, Interoperable, and Reusable) and FACT (Fair, Accurate, Confidential, and Transparent) principles. For additional information, please visit \url{www.sobigdata.eu}.}
To aid reproducibility we provide the data and source code needed to replicate our findings~\cite{replication}.

%% file: main.bbl

\begin{thebibliography}{47}

%%% the Web address.

\ifx \showCODEN    \undefined \def \showCODEN     #1{\unskip}     \fi
\ifx \showDOI      \undefined \def \showDOI       #1{#1}\fi
\ifx \showISBNx    \undefined \def \showISBNx     #1{\unskip}     \fi
\ifx \showISBNxiii \undefined \def \showISBNxiii  #1{\unskip}     \fi
\ifx \showISSN     \undefined \def \showISSN      #1{\unskip}     \fi
\ifx \showLCCN     \undefined \def \showLCCN      #1{\unskip}     \fi
\ifx \shownote     \undefined \def \shownote      #1{#1}          \fi
\ifx \showarticletitle \undefined \def \showarticletitle #1{#1}   \fi
\ifx \showURL      \undefined \def \showURL       {\relax}        \fi
% invisible to TeX
\providecommand\bibfield[2]{#2}
\providecommand\bibinfo[2]{#2}
\providecommand\natexlab[1]{#1}
\providecommand\showeprint[2][]{arXiv:#2}

\bibitem[Ahmed et~al\mbox{.}(2016)]%
        {ahmed2016}
\bibfield{author}{\bibinfo{person}{Tarek~M. Ahmed}, \bibinfo{person}{Cor-Paul Bezemer}, \bibinfo{person}{Tse-Hsun Chen}, \bibinfo{person}{Ahmed~E. Hassan}, {and} \bibinfo{person}{Weiyi Shang}.} \bibinfo{year}{2016}\natexlab{}.
\newblock \showarticletitle{Studying the Effectiveness of Application Performance Management (APM) Tools for Detecting Performance Regressions for Web Applications: An Experience Report}. In \bibinfo{booktitle}{\emph{Proceedings of the 13th International Conference on Mining Software Repositories}} (Austin, Texas) \emph{(\bibinfo{series}{MSR '16})}. \bibinfo{publisher}{Association for Computing Machinery}, \bibinfo{address}{New York, NY, USA}, \bibinfo{pages}{1–12}.
\newblock
\showISBNx{9781450341868}
\urldef\tempurl%
\url{https://doi.org/10.1145/2901739.2901774}
\showDOI{\tempurl}


\bibitem[Anand et~al\mbox{.}(2020)]%
        {anand2020}
\bibfield{author}{\bibinfo{person}{Vaastav Anand}, \bibinfo{person}{Matheus Stolet}, \bibinfo{person}{Thomas Davidson}, \bibinfo{person}{Ivan Beschastnikh}, \bibinfo{person}{Tamara Munzner}, {and} \bibinfo{person}{Jonathan Mace}.} \bibinfo{year}{2020}\natexlab{}.
\newblock \bibinfo{title}{Aggregate-{Driven} {Trace} {Visualizations} for {Performance} {Debugging}}.
\newblock
\newblock
\newblock
\shownote{arXiv: 2010.13681 [cs] Number: arXiv:2010.13681}.


\bibitem[Ardelean et~al\mbox{.}(2018)]%
        {ardelean2018}
\bibfield{author}{\bibinfo{person}{Dan Ardelean}, \bibinfo{person}{Amer Diwan}, {and} \bibinfo{person}{Chandra Erdman}.} \bibinfo{year}{2018}\natexlab{}.
\newblock \showarticletitle{Performance {Analysis} of {Cloud} {Applications}}. In \bibinfo{booktitle}{\emph{Proceedings of the 15th {USENIX} {Conference} on {Networked} {Systems} {Design} and {Implementation}}} \emph{(\bibinfo{series}{{NSDI}'18})}. \bibinfo{publisher}{USENIX Association}, \bibinfo{address}{USA}, \bibinfo{pages}{405--417}.
\newblock
\showISBNx{978-1-931971-43-0}
\newblock
\shownote{event-place: Renton, WA, USA}.


\bibitem[Avritzer et~al\mbox{.}(2019)]%
        {avritzer2019}
\bibfield{author}{\bibinfo{person}{Alberto Avritzer}, \bibinfo{person}{Daniel Menasch\'{e}}, \bibinfo{person}{Vilc Rufino}, \bibinfo{person}{Barbara Russo}, \bibinfo{person}{Andrea Janes}, \bibinfo{person}{Vincenzo Ferme}, \bibinfo{person}{Andr\'{e} van Hoorn}, {and} \bibinfo{person}{Henning Schulz}.} \bibinfo{year}{2019}\natexlab{}.
\newblock \showarticletitle{PPTAM: Production and Performance Testing Based Application Monitoring}. In \bibinfo{booktitle}{\emph{Companion of the 2019 ACM/SPEC International Conference on Performance Engineering}} (Mumbai, India) \emph{(\bibinfo{series}{ICPE '19})}. \bibinfo{publisher}{Association for Computing Machinery}, \bibinfo{address}{New York, NY, USA}, \bibinfo{pages}{39–40}.
\newblock
\showISBNx{9781450362863}
\urldef\tempurl%
\url{https://doi.org/10.1145/3302541.3311961}
\showDOI{\tempurl}


\bibitem[Beck et~al\mbox{.}(2021)]%
        {beck2021}
\bibfield{author}{\bibinfo{person}{Samuel Beck}, \bibinfo{person}{Sebastian Frank}, \bibinfo{person}{Alireza Hakamian}, \bibinfo{person}{Leonel Merino}, {and} \bibinfo{person}{Andr{\'e} van Hoorn}.} \bibinfo{year}{2021}\natexlab{}.
\newblock \showarticletitle{TransVis: Using visualizations and chatbots for supporting transient behavior in microservice systems}. In \bibinfo{booktitle}{\emph{2021 Working Conference on Software Visualization (VISSOFT)}}. IEEE, \bibinfo{pages}{65--75}.
\newblock


\bibitem[Beschastnikh et~al\mbox{.}(2020)]%
        {beschastnikh2020}
\bibfield{author}{\bibinfo{person}{Ivan Beschastnikh}, \bibinfo{person}{Perry Liu}, \bibinfo{person}{Albert Xing}, \bibinfo{person}{Patty Wang}, \bibinfo{person}{Yuriy Brun}, {and} \bibinfo{person}{Michael~D. Ernst}.} \bibinfo{year}{2020}\natexlab{}.
\newblock \showarticletitle{Visualizing {Distributed} {System} {Executions}}.
\newblock \bibinfo{journal}{\emph{ACM Trans. Softw. Eng. Methodol.}} \bibinfo{volume}{29}, \bibinfo{number}{2} (\bibinfo{date}{March} \bibinfo{year}{2020}).
\newblock
\showISSN{1049-331X}
\urldef\tempurl%
\url{https://doi.org/10.1145/3375633}
\showDOI{\tempurl}
\newblock
\shownote{Place: New York, NY, USA Publisher: Association for Computing Machinery}.


\bibitem[Boyd and Vandenberghe(2004)]%
        {Boyd2004}
\bibfield{author}{\bibinfo{person}{Stephen Boyd} {and} \bibinfo{person}{Lieven Vandenberghe}.} \bibinfo{year}{2004}\natexlab{}.
\newblock \bibinfo{booktitle}{\emph{Convex Optimization}}.
\newblock \bibinfo{publisher}{Cambridge University Press}.
\newblock
\urldef\tempurl%
\url{https://doi.org/10.1017/CBO9780511804441}
\showDOI{\tempurl}


\bibitem[Cortellessa and Traini(2020)]%
        {cortellessa2020}
\bibfield{author}{\bibinfo{person}{Vittorio Cortellessa} {and} \bibinfo{person}{Luca Traini}.} \bibinfo{year}{2020}\natexlab{}.
\newblock \showarticletitle{Detecting {Latency} {Degradation} {Patterns} in {Service}-{Based} {Systems}}. In \bibinfo{booktitle}{\emph{Proceedings of the {ACM}/{SPEC} {International} {Conference} on {Performance} {Engineering}}} \emph{(\bibinfo{series}{{ICPE} '20})}. \bibinfo{publisher}{Association for Computing Machinery}, \bibinfo{address}{New York, NY, USA}, \bibinfo{pages}{161--172}.
\newblock
\showISBNx{978-1-4503-6991-6}
\urldef\tempurl%
\url{https://doi.org/10.1145/3358960.3379126}
\showDOI{\tempurl}


\bibitem[Davidson and Mace(2022)]%
        {davidson2022}
\bibfield{author}{\bibinfo{person}{Thomas Davidson} {and} \bibinfo{person}{Jonathan Mace}.} \bibinfo{year}{2022}\natexlab{}.
\newblock \showarticletitle{See it to believe it? {The} role of visualisation in systems research}. In \bibinfo{booktitle}{\emph{Proceedings of the 13th {Symposium} on {Cloud} {Computing}}} \emph{(\bibinfo{series}{{SoCC} '22})}. \bibinfo{publisher}{Association for Computing Machinery}, \bibinfo{address}{New York, NY, USA}, \bibinfo{pages}{419--428}.
\newblock
\showISBNx{978-1-4503-9414-7}
\urldef\tempurl%
\url{https://doi.org/10.1145/3542929.3563488}
\showDOI{\tempurl}


\bibitem[Davidson et~al\mbox{.}(2023)]%
        {Davidson2023}
\bibfield{author}{\bibinfo{person}{Thomas Davidson}, \bibinfo{person}{Emily Wall}, {and} \bibinfo{person}{Jonathan Mace}.} \bibinfo{year}{2023}\natexlab{}.
\newblock \showarticletitle{A Qualitative Interview Study of Distributed Tracing Visualisation: A Characterisation of Challenges and Opportunities}.
\newblock \bibinfo{journal}{\emph{IEEE Transactions on Visualization and Computer Graphics}} (\bibinfo{year}{2023}), \bibinfo{pages}{1--12}.
\newblock
\urldef\tempurl%
\url{https://doi.org/10.1109/TVCG.2023.3241596}
\showDOI{\tempurl}


\bibitem[Dean and Barroso(2013)]%
        {Dean2013}
\bibfield{author}{\bibinfo{person}{Jeffrey Dean} {and} \bibinfo{person}{Luiz~André Barroso}.} \bibinfo{year}{2013}\natexlab{}.
\newblock \showarticletitle{The Tail at Scale}.
\newblock \bibinfo{journal}{\emph{Commun. ACM}}  \bibinfo{volume}{56} (\bibinfo{year}{2013}), \bibinfo{pages}{74--80}.
\newblock
\urldef\tempurl%
\url{http://cacm.acm.org/magazines/2013/2/160173-the-tail-at-scale/fulltext}
\showURL{%
\tempurl}


\bibitem[Elastic(2023a)]%
        {elasticsearch}
\bibfield{author}{\bibinfo{person}{Elastic}.} \bibinfo{year}{2023}\natexlab{a}.
\newblock \bibinfo{title}{Elasticsearch: The Official Distributed Search \& Analytics Engine}.
\newblock
\newblock
\urldef\tempurl%
\url{https://www.elastic.co/elasticsearch/}
\showURL{%
\tempurl}
\newblock
\shownote{"Accessed 2023-01-15 18:38"}.


\bibitem[Elastic(2023b)]%
        {kibana}
\bibfield{author}{\bibinfo{person}{Elastic}.} \bibinfo{year}{2023}\natexlab{b}.
\newblock \bibinfo{title}{Kibana: {Explore}, {Visualize}, {Discover} {Data}}.
\newblock
\newblock
\urldef\tempurl%
\url{https://www.elastic.co/kibana}
\showURL{%
\tempurl}
\newblock
\shownote{"Accessed 2023-01-15 17:38"}.


\bibitem[Everitt(1998)]%
        {everitt1998}
\bibfield{author}{\bibinfo{person}{Brian Everitt}.} \bibinfo{year}{1998}\natexlab{}.
\newblock \bibinfo{booktitle}{\emph{The {Cambridge} {Dictionary} of {Statistics}}}.
\newblock \bibinfo{publisher}{Cambridge University Press}, \bibinfo{address}{Cambridge, UK}.
\newblock
\showISBNx{978-0521593465}


\bibitem[Farro(2018)]%
        {farro2018}
\bibfield{author}{\bibinfo{person}{Joe Farro}.} \bibinfo{year}{2018}\natexlab{}.
\newblock \bibinfo{title}{Trace comparisons arrive in {Jaeger} 1.7}.
\newblock
\newblock
\urldef\tempurl%
\url{https://medium.com/jaegertracing/trace-comparisons-arrive-in-jaeger-1-7-a97ad5e2d05d}
\showURL{%
\tempurl}
\newblock
\shownote{"Accessed 2023-01-15 18:16"}.


\bibitem[Inc.(2023)]%
        {dynatraceflow}
\bibfield{author}{\bibinfo{person}{Dynatrace Inc.}} \bibinfo{year}{2023}\natexlab{}.
\newblock \bibinfo{title}{Dynatrace. Service Flow}.
\newblock
\newblock
\urldef\tempurl%
\url{https://www.dynatrace.com/platform/service-flow/}
\showURL{%
\tempurl}
\newblock
\shownote{"Accessed 2023-02-14 14:10:59"}.


\bibitem[Jiang and Hassan(2015)]%
        {jiang2015}
\bibfield{author}{\bibinfo{person}{Zhen~Ming Jiang} {and} \bibinfo{person}{Ahmed~E. Hassan}.} \bibinfo{year}{2015}\natexlab{}.
\newblock \showarticletitle{A Survey on Load Testing of Large-Scale Software Systems}.
\newblock \bibinfo{journal}{\emph{IEEE Transactions on Software Engineering}} \bibinfo{volume}{41}, \bibinfo{number}{11} (\bibinfo{year}{2015}), \bibinfo{pages}{1091--1118}.
\newblock
\urldef\tempurl%
\url{https://doi.org/10.1109/TSE.2015.2445340}
\showDOI{\tempurl}


\bibitem[Jonatan et~al\mbox{.}(2023)]%
        {locust}
\bibfield{author}{\bibinfo{person}{Heyman Jonatan}, \bibinfo{person}{Carl Byström}, \bibinfo{person}{Joakim Hamrén}, {and} \bibinfo{person}{Hugo Heyman}.} \bibinfo{year}{2023}\natexlab{}.
\newblock \bibinfo{title}{An open source load testing tool.}
\newblock
\newblock
\urldef\tempurl%
\url{https://locust.io}
\showURL{%
\tempurl}
\newblock
\shownote{"Accessed 2023-06-10 13:38"}.


\bibitem[Kaldor et~al\mbox{.}(2017)]%
        {kaldor2017}
\bibfield{author}{\bibinfo{person}{Jonathan Kaldor}, \bibinfo{person}{Jonathan Mace}, \bibinfo{person}{Micha{\textbackslash}l Bejda}, \bibinfo{person}{Edison Gao}, \bibinfo{person}{Wiktor Kuropatwa}, \bibinfo{person}{Joe O'Neill}, \bibinfo{person}{Kian~Win Ong}, \bibinfo{person}{Bill Schaller}, \bibinfo{person}{Pingjia Shan}, \bibinfo{person}{Brendan Viscomi}, \bibinfo{person}{Vinod Venkataraman}, \bibinfo{person}{Kaushik Veeraraghavan}, {and} \bibinfo{person}{Yee~Jiun Song}.} \bibinfo{year}{2017}\natexlab{}.
\newblock \showarticletitle{Canopy: {An} {End}-to-{End} {Performance} {Tracing} {And} {Analysis} {System}}. In \bibinfo{booktitle}{\emph{Proceedings of the 26th {Symposium} on {Operating} {Systems} {Principles}}} \emph{(\bibinfo{series}{{SOSP} '17})}. \bibinfo{publisher}{Association for Computing Machinery}, \bibinfo{address}{New York, NY, USA}, \bibinfo{pages}{34--50}.
\newblock
\showISBNx{978-1-4503-5085-3}
\urldef\tempurl%
\url{https://doi.org/10.1145/3132747.3132749}
\showDOI{\tempurl}


\bibitem[Krushevskaja and Sandler(2013)]%
        {krushevskaja2013}
\bibfield{author}{\bibinfo{person}{Darja Krushevskaja} {and} \bibinfo{person}{Mark Sandler}.} \bibinfo{year}{2013}\natexlab{}.
\newblock \showarticletitle{Understanding Latency Variations of Black Box Services}. In \bibinfo{booktitle}{\emph{Proceedings of the 22nd International Conference on World Wide Web}} (Rio de Janeiro, Brazil) \emph{(\bibinfo{series}{WWW '13})}. \bibinfo{publisher}{Association for Computing Machinery}, \bibinfo{address}{New York, NY, USA}, \bibinfo{pages}{703–714}.
\newblock
\showISBNx{9781450320351}
\urldef\tempurl%
\url{https://doi.org/10.1145/2488388.2488450}
\showDOI{\tempurl}


\bibitem[Laaber and Leitner(2018)]%
        {laaber2018}
\bibfield{author}{\bibinfo{person}{Christoph Laaber} {and} \bibinfo{person}{Philipp Leitner}.} \bibinfo{year}{2018}\natexlab{}.
\newblock \showarticletitle{An Evaluation of Open-Source Software Microbenchmark Suites for Continuous Performance Assessment}. In \bibinfo{booktitle}{\emph{Proceedings of the 15th International Conference on Mining Software Repositories}} (Gothenburg, Sweden) \emph{(\bibinfo{series}{MSR '18})}. \bibinfo{publisher}{Association for Computing Machinery}, \bibinfo{address}{New York, NY, USA}, \bibinfo{pages}{119–130}.
\newblock
\showISBNx{9781450357166}
\urldef\tempurl%
\url{https://doi.org/10.1145/3196398.3196407}
\showDOI{\tempurl}


\bibitem[Leone and Traini(2023)]%
        {Leone2023}
\bibfield{author}{\bibinfo{person}{Jessica Leone} {and} \bibinfo{person}{Luca Traini}.} \bibinfo{year}{2023}\natexlab{}.
\newblock \showarticletitle{Enhancing Trace Visualizations for Microservices Performance Analysis}. In \bibinfo{booktitle}{\emph{Companion of the 2023 ACM/SPEC International Conference on Performance Engineering}} (Coimbra, Portugal) \emph{(\bibinfo{series}{ICPE '23 Companion})}. \bibinfo{publisher}{Association for Computing Machinery}, \bibinfo{address}{New York, NY, USA}, \bibinfo{pages}{283–287}.
\newblock
\showISBNx{9798400700729}
\urldef\tempurl%
\url{https://doi.org/10.1145/3578245.3584729}
\showDOI{\tempurl}


\bibitem[Li et~al\mbox{.}(2021)]%
        {li2022}
\bibfield{author}{\bibinfo{person}{Bowen Li}, \bibinfo{person}{Xin Peng}, \bibinfo{person}{Qilin Xiang}, \bibinfo{person}{Hanzhang Wang}, \bibinfo{person}{Tao Xie}, \bibinfo{person}{Jun Sun}, {and} \bibinfo{person}{Xuanzhe Liu}.} \bibinfo{year}{2021}\natexlab{}.
\newblock \showarticletitle{Enjoy your observability: an industrial survey of microservice tracing and analysis}.
\newblock \bibinfo{journal}{\emph{Empirical Software Engineering}} \bibinfo{volume}{27}, \bibinfo{number}{1} (\bibinfo{year}{2021}), \bibinfo{pages}{25}.
\newblock
\showISBNx{1573-7616}
\urldef\tempurl%
\url{https://doi.org/10.1007/s10664-021-10063-9}
\showDOI{\tempurl}


\bibitem[Liao et~al\mbox{.}(2020)]%
        {liao2020}
\bibfield{author}{\bibinfo{person}{Lizhi Liao}, \bibinfo{person}{Jinfu Chen}, \bibinfo{person}{Heng Li}, \bibinfo{person}{Yi Zeng}, \bibinfo{person}{Weiyi Shang}, \bibinfo{person}{Jianmei Guo}, \bibinfo{person}{Catalin Sporea}, \bibinfo{person}{Andrei Toma}, {and} \bibinfo{person}{Sarah Sajedi}.} \bibinfo{year}{2020}\natexlab{}.
\newblock \showarticletitle{Using Black-Box Performance Models to Detect Performance Regressions under Varying Workloads: An Empirical Study}.
\newblock \bibinfo{journal}{\emph{Empirical Softw. Engg.}} \bibinfo{volume}{25}, \bibinfo{number}{5} (\bibinfo{date}{sep} \bibinfo{year}{2020}), \bibinfo{pages}{4130–4160}.
\newblock
\showISSN{1382-3256}
\urldef\tempurl%
\url{https://doi.org/10.1007/s10664-020-09866-z}
\showDOI{\tempurl}


\bibitem[Liao et~al\mbox{.}(2021)]%
        {liao2021}
\bibfield{author}{\bibinfo{person}{Lizhi Liao}, \bibinfo{person}{Jinfu Chen}, \bibinfo{person}{Heng Li}, \bibinfo{person}{Yi Zeng}, \bibinfo{person}{Weiyi Shang}, \bibinfo{person}{Catalin Sporea}, \bibinfo{person}{Andrei Toma}, {and} \bibinfo{person}{Sarah Sajedi}.} \bibinfo{year}{2021}\natexlab{}.
\newblock \showarticletitle{Locating Performance Regression Root Causes in the Field Operations of Web-based Systems: An Experience Report}.
\newblock \bibinfo{journal}{\emph{IEEE Transactions on Software Engineering}} (\bibinfo{year}{2021}), \bibinfo{pages}{1--1}.
\newblock
\urldef\tempurl%
\url{https://doi.org/10.1109/TSE.2021.3131529}
\showDOI{\tempurl}


\bibitem[Lloyd(1982)]%
        {Lloyd1982}
\bibfield{author}{\bibinfo{person}{S. Lloyd}.} \bibinfo{year}{1982}\natexlab{}.
\newblock \showarticletitle{Least squares quantization in PCM}.
\newblock \bibinfo{journal}{\emph{IEEE Transactions on Information Theory}} \bibinfo{volume}{28}, \bibinfo{number}{2} (\bibinfo{year}{1982}), \bibinfo{pages}{129--137}.
\newblock
\urldef\tempurl%
\url{https://doi.org/10.1109/TIT.1982.1056489}
\showDOI{\tempurl}


\bibitem[Luo et~al\mbox{.}(2021)]%
        {luo2021}
\bibfield{author}{\bibinfo{person}{Shutian Luo}, \bibinfo{person}{Huanle Xu}, \bibinfo{person}{Chengzhi Lu}, \bibinfo{person}{Kejiang Ye}, \bibinfo{person}{Guoyao Xu}, \bibinfo{person}{Liping Zhang}, \bibinfo{person}{Yu Ding}, \bibinfo{person}{Jian He}, {and} \bibinfo{person}{Chengzhong Xu}.} \bibinfo{year}{2021}\natexlab{}.
\newblock \showarticletitle{Characterizing Microservice Dependency and Performance: Alibaba Trace Analysis}. In \bibinfo{booktitle}{\emph{Proceedings of the ACM Symposium on Cloud Computing}}. \bibinfo{pages}{412--426}.
\newblock


\bibitem[Mace(2017)]%
        {mace2017}
\bibfield{author}{\bibinfo{person}{Jonathan Mace}.} \bibinfo{year}{2017}\natexlab{}.
\newblock \bibinfo{booktitle}{\emph{{End-to-End Tracing: Adoption and Use Cases}}}.
\newblock \bibinfo{type}{{Survey}}. \bibinfo{institution}{Brown University}.
\newblock
\urldef\tempurl%
\url{https://cs.brown.edu/people/jcmace/papers/mace2017survey.pdf}
\showURL{%
\tempurl}


\bibitem[Majors et~al\mbox{.}(2022)]%
        {majors2022}
\bibfield{author}{\bibinfo{person}{C. Majors}, \bibinfo{person}{L. Fong-Jones}, {and} \bibinfo{person}{G. Miranda}.} \bibinfo{year}{2022}\natexlab{}.
\newblock \bibinfo{booktitle}{\emph{Observability Engineering: Achieving Production Excellence}}.
\newblock \bibinfo{publisher}{O'Reilly Media, Incorporated}.
\newblock
\showISBNx{9781492076445}
\showLCCN{2023275144}
\urldef\tempurl%
\url{https://books.google.it/books?id=MbmLzgEACAAJ}
\showURL{%
\tempurl}


\bibitem[Newman(2015)]%
        {newman2015}
\bibfield{author}{\bibinfo{person}{Sam Newman}.} \bibinfo{year}{2015}\natexlab{}.
\newblock \bibinfo{booktitle}{\emph{Building {Microservices}} (\bibinfo{edition}{1st} ed.)}.
\newblock \bibinfo{publisher}{O'Reilly Media, Inc.}
\newblock
\showISBNx{978-1-4919-5035-7}


\bibitem[O'Hanlon(2006)]%
        {ohanlon2006}
\bibfield{author}{\bibinfo{person}{Charlene O'Hanlon}.} \bibinfo{year}{2006}\natexlab{}.
\newblock \showarticletitle{A {Conversation} with {Werner} {Vogels}}.
\newblock \bibinfo{journal}{\emph{Queue}} \bibinfo{volume}{4}, \bibinfo{number}{4} (\bibinfo{date}{May} \bibinfo{year}{2006}), \bibinfo{pages}{14:14--14:22}.
\newblock
\showISSN{1542-7730}
\urldef\tempurl%
\url{https://doi.org/10.1145/1142055.1142065}
\showDOI{\tempurl}
\newblock
\shownote{Place: New York, NY, USA Publisher: ACM}.


\bibitem[Parker et~al\mbox{.}(2020)]%
        {parker2020}
\bibfield{author}{\bibinfo{person}{Austin Parker}, \bibinfo{person}{Daniel Spoonhower}, \bibinfo{person}{Jonathan Mace}, \bibinfo{person}{Ben Sigelman}, {and} \bibinfo{person}{Rebecca Isaacs}.} \bibinfo{year}{2020}\natexlab{}.
\newblock \bibinfo{booktitle}{\emph{Distributed {Tracing} in {Practice}: {Instrumenting}, {Analyzing}, and {Debugging} {Microservices}}}.
\newblock \bibinfo{publisher}{O'Reilly Media, Incorporated}.
\newblock
\showISBNx{978-1-4920-5663-8}


\bibitem[Rousseeuw(1987)]%
        {Rousseeuw1987}
\bibfield{author}{\bibinfo{person}{Peter~J. Rousseeuw}.} \bibinfo{year}{1987}\natexlab{}.
\newblock \showarticletitle{Silhouettes: A graphical aid to the interpretation and validation of cluster analysis}.
\newblock \bibinfo{journal}{\emph{J. Comput. Appl. Math.}}  \bibinfo{volume}{20} (\bibinfo{year}{1987}), \bibinfo{pages}{53--65}.
\newblock
\showISSN{0377-0427}
\urldef\tempurl%
\url{https://doi.org/10.1016/0377-0427(87)90125-7}
\showDOI{\tempurl}


\bibitem[Rubin and Rinard(2016)]%
        {rubin2016}
\bibfield{author}{\bibinfo{person}{Julia Rubin} {and} \bibinfo{person}{Martin Rinard}.} \bibinfo{year}{2016}\natexlab{}.
\newblock \showarticletitle{The {Challenges} of {Staying} {Together} {While} {Moving} {Fast}: {An} {Exploratory} {Study}}. In \bibinfo{booktitle}{\emph{Proceedings of the 38th {International} {Conference} on {Software} {Engineering}}} \emph{(\bibinfo{series}{{ICSE} '16})}. \bibinfo{publisher}{Association for Computing Machinery}, \bibinfo{address}{New York, NY, USA}, \bibinfo{pages}{982--993}.
\newblock
\showISBNx{978-1-4503-3900-1}
\urldef\tempurl%
\url{https://doi.org/10.1145/2884781.2884871}
\showDOI{\tempurl}


\bibitem[Sambasivan et~al\mbox{.}(2016)]%
        {sambasivan2016}
\bibfield{author}{\bibinfo{person}{Raja~R. Sambasivan}, \bibinfo{person}{Ilari Shafer}, \bibinfo{person}{Jonathan Mace}, \bibinfo{person}{Benjamin~H. Sigelman}, \bibinfo{person}{Rodrigo Fonseca}, {and} \bibinfo{person}{Gregory~R. Ganger}.} \bibinfo{year}{2016}\natexlab{}.
\newblock \showarticletitle{Principled {Workflow}-centric {Tracing} of {Distributed} {Systems}}. In \bibinfo{booktitle}{\emph{Proceedings of the {Seventh} {ACM} {Symposium} on {Cloud} {Computing}}} \emph{(\bibinfo{series}{{SoCC} '16})}. \bibinfo{publisher}{ACM}, \bibinfo{address}{New York, NY, USA}, \bibinfo{pages}{401--414}.
\newblock
\showISBNx{978-1-4503-4525-5}
\urldef\tempurl%
\url{https://doi.org/10.1145/2987550.2987568}
\showDOI{\tempurl}


\bibitem[Sambasivan et~al\mbox{.}(2013)]%
        {sambasivan2013}
\bibfield{author}{\bibinfo{person}{R.~R. Sambasivan}, \bibinfo{person}{I. Shafer}, \bibinfo{person}{M.~L. Mazurek}, {and} \bibinfo{person}{G.~R. Ganger}.} \bibinfo{year}{2013}\natexlab{}.
\newblock \showarticletitle{Visualizing {Request}-{Flow} {Comparison} to {Aid} {Performance} {Diagnosis} in {Distributed} {Systems}}.
\newblock \bibinfo{journal}{\emph{IEEE Transactions on Visualization and Computer Graphics}} \bibinfo{volume}{19}, \bibinfo{number}{12} (\bibinfo{year}{2013}), \bibinfo{pages}{2466--2475}.
\newblock
\urldef\tempurl%
\url{https://doi.org/10.1109/TVCG.2013.233}
\showDOI{\tempurl}


\bibitem[Sigelman et~al\mbox{.}(2010)]%
        {sigelman2010}
\bibfield{author}{\bibinfo{person}{Benjamin~H. Sigelman}, \bibinfo{person}{Luiz~André Barroso}, \bibinfo{person}{Mike Burrows}, \bibinfo{person}{Pat Stephenson}, \bibinfo{person}{Manoj Plakal}, \bibinfo{person}{Donald Beaver}, \bibinfo{person}{Saul Jaspan}, {and} \bibinfo{person}{Chandan Shanbhag}.} \bibinfo{year}{2010}\natexlab{}.
\newblock \bibinfo{booktitle}{\emph{Dapper, a {Large}-{Scale} {Distributed} {Systems} {Tracing} {Infrastructure}}}.
\newblock \bibinfo{type}{{T}echnical {R}eport}. \bibinfo{institution}{Google}.
\newblock


\bibitem[Sridharan(2017)]%
        {sridharan2017}
\bibfield{author}{\bibinfo{person}{Cindy Sridharan}.} \bibinfo{year}{2017}\natexlab{}.
\newblock \bibinfo{title}{Testing Microservices, the sane way}.
\newblock
\newblock
\urldef\tempurl%
\url{https://copyconstruct.medium.com/testing-microservices-the-sane-way-9bb31d158c16}
\showURL{%
\tempurl}
\newblock
\shownote{"Accessed 2023-01-16 11:14"}.


\bibitem[Technologies(2023)]%
        {jaeger}
\bibfield{author}{\bibinfo{person}{Uber Technologies}.} \bibinfo{year}{2023}\natexlab{}.
\newblock \bibinfo{title}{Jaeger: {Open} source, end-to-end distributed tracing}.
\newblock
\newblock
\urldef\tempurl%
\url{https://www.jaegertracing.io/}
\showURL{%
\tempurl}
\newblock
\shownote{"Accessed 2023-01-15 14:10:59"}.


\bibitem[Traini(2022)]%
        {traini2022}
\bibfield{author}{\bibinfo{person}{Luca Traini}.} \bibinfo{year}{2022}\natexlab{}.
\newblock \showarticletitle{Exploring {Performance} {Assurance} {Practices} and {Challenges} in {Agile} {Software} {Development}: {An} {Ethnographic} {Study}}.
\newblock \bibinfo{journal}{\emph{Empirical Software Engineering}} \bibinfo{volume}{27}, \bibinfo{number}{3} (\bibinfo{year}{2022}), \bibinfo{pages}{74}.
\newblock
\urldef\tempurl%
\url{https://doi.org/10.1007/s10664-021-10069-3}
\showDOI{\tempurl}
\newblock
\shownote{ISBN: 1573-7616}.


\bibitem[Traini and Cortellessa(2023)]%
        {traini2022b}
\bibfield{author}{\bibinfo{person}{Luca Traini} {and} \bibinfo{person}{Vittorio Cortellessa}.} \bibinfo{year}{2023}\natexlab{}.
\newblock \showarticletitle{DeLag: Using Multi-Objective Optimization to Enhance the Detection of Latency Degradation Patterns in Service-Based Systems}.
\newblock \bibinfo{journal}{\emph{IEEE Transactions on Software Engineering}} (\bibinfo{year}{2023}), \bibinfo{pages}{1--28}.
\newblock
\urldef\tempurl%
\url{https://doi.org/10.1109/TSE.2023.3266041}
\showDOI{\tempurl}


\bibitem[Traini et~al\mbox{.}(2022)]%
        {traini2022c}
\bibfield{author}{\bibinfo{person}{Luca Traini}, \bibinfo{person}{Vittorio Cortellessa}, \bibinfo{person}{Daniele Di~Pompeo}, {and} \bibinfo{person}{Michele Tucci}.} \bibinfo{year}{2022}\natexlab{}.
\newblock \showarticletitle{Towards effective assessment of steady state performance in Java software: are we there yet?}
\newblock \bibinfo{journal}{\emph{Empirical Software Engineering}} \bibinfo{volume}{28}, \bibinfo{number}{1} (\bibinfo{year}{2022}), \bibinfo{pages}{13}.
\newblock
\showISBNx{1573-7616}
\urldef\tempurl%
\url{https://doi.org/10.1007/s10664-022-10247-x}
\showDOI{\tempurl}


\bibitem[Traini et~al\mbox{.}(2021)]%
        {traini2021}
\bibfield{author}{\bibinfo{person}{Luca Traini}, \bibinfo{person}{Daniele Di~Pompeo}, \bibinfo{person}{Michele Tucci}, \bibinfo{person}{Bin Lin}, \bibinfo{person}{Simone Scalabrino}, \bibinfo{person}{Gabriele Bavota}, \bibinfo{person}{Michele Lanza}, \bibinfo{person}{Rocco Oliveto}, {and} \bibinfo{person}{Vittorio Cortellessa}.} \bibinfo{year}{2021}\natexlab{}.
\newblock \showarticletitle{How Software Refactoring Impacts Execution Time}.
\newblock \bibinfo{journal}{\emph{ACM Trans. Softw. Eng. Methodol.}} \bibinfo{volume}{31}, \bibinfo{number}{2}, Article \bibinfo{articleno}{25} (\bibinfo{date}{dec} \bibinfo{year}{2021}), \bibinfo{numpages}{23}~pages.
\newblock
\showISSN{1049-331X}
\urldef\tempurl%
\url{https://doi.org/10.1145/3485136}
\showDOI{\tempurl}


\bibitem[Traini et~al\mbox{.}(2023)]%
        {replication}
\bibfield{author}{\bibinfo{person}{Luca Traini}, \bibinfo{person}{Jessica Leone}, \bibinfo{person}{Giovanni Stilo}, {and} \bibinfo{person}{Antinisca Di~Marco}.} \bibinfo{year}{2023}\natexlab{}.
\newblock \bibinfo{title}{VAMP - Replication Package}.
\newblock \bibinfo{howpublished}{\url{https://github.com/lucatraini/VAMP}}.
\newblock


\bibitem[Veeraraghavan et~al\mbox{.}(2016)]%
        {veeraraghavan2016}
\bibfield{author}{\bibinfo{person}{Kaushik Veeraraghavan}, \bibinfo{person}{Justin Meza}, \bibinfo{person}{David Chou}, \bibinfo{person}{Wonho Kim}, \bibinfo{person}{Sonia Margulis}, \bibinfo{person}{Scott Michelson}, \bibinfo{person}{Rajesh Nishtala}, \bibinfo{person}{Daniel Obenshain}, \bibinfo{person}{Dmitri Perelman}, {and} \bibinfo{person}{Yee~Jiun Song}.} \bibinfo{year}{2016}\natexlab{}.
\newblock \showarticletitle{Kraken: {Leveraging} {Live} {Traffic} {Tests} to {Identify} and {Resolve} {Resource} {Utilization} {Bottlenecks} in {Large} {Scale} {Web} {Services}}. In \bibinfo{booktitle}{\emph{12th {USENIX} {Symposium} on {Operating} {Systems} {Design} and {Implementation} ({OSDI} 16)}}. \bibinfo{publisher}{USENIX Association}, \bibinfo{address}{Savannah, GA}, \bibinfo{pages}{635--651}.
\newblock
\showISBNx{978-1-931971-33-1}


\bibitem[Zhang et~al\mbox{.}(2022)]%
        {zhang2022}
\bibfield{author}{\bibinfo{person}{Chenxi Zhang}, \bibinfo{person}{Xin Peng}, \bibinfo{person}{Chaofeng Sha}, \bibinfo{person}{Ke Zhang}, \bibinfo{person}{Zhenqing Fu}, \bibinfo{person}{Xiya Wu}, \bibinfo{person}{Qingwei Lin}, {and} \bibinfo{person}{Dongmei Zhang}.} \bibinfo{year}{2022}\natexlab{}.
\newblock \showarticletitle{DeepTraLog: Trace-Log Combined Microservice Anomaly Detection through Graph-based Deep Learning}. In \bibinfo{booktitle}{\emph{2022 IEEE/ACM 44th International Conference on Software Engineering (ICSE)}}. \bibinfo{pages}{623--634}.
\newblock
\urldef\tempurl%
\url{https://doi.org/10.1145/3510003.3510180}
\showDOI{\tempurl}


\bibitem[Zhou et~al\mbox{.}(2021)]%
        {zhou2021}
\bibfield{author}{\bibinfo{person}{Xiang Zhou}, \bibinfo{person}{Xin Peng}, \bibinfo{person}{Tao Xie}, \bibinfo{person}{Jun Sun}, \bibinfo{person}{Chao Ji}, \bibinfo{person}{Wenhai Li}, {and} \bibinfo{person}{Dan Ding}.} \bibinfo{year}{2021}\natexlab{}.
\newblock \showarticletitle{Fault Analysis and Debugging of Microservice Systems: Industrial Survey, Benchmark System, and Empirical Study}.
\newblock \bibinfo{journal}{\emph{IEEE Transactions on Software Engineering}} \bibinfo{volume}{47}, \bibinfo{number}{2} (\bibinfo{year}{2021}), \bibinfo{pages}{243--260}.
\newblock
\urldef\tempurl%
\url{https://doi.org/10.1109/TSE.2018.2887384}
\showDOI{\tempurl}


\end{thebibliography}
